\documentclass[aps,pra,preprint,superscriptaddress]{revtex4-1}

\usepackage{graphicx}
\usepackage{dcolumn}
\usepackage{bm}
\begin{document}


\title{Propagation of First and Second Sound in a Highly-Elongated Trapped Bose Condensed Gas at Finite temperatures}


\author{Emiko Arahata}
\email{arahata@vortex.c.u-tokyo.ac.jp}
\affiliation{Department of Basic Science, The University of Tokyo, 3-8-1  Komaba, Meguro-ku, Tokyo,  153-8902, Japan}
\author{Tetsuro Nikuni}
\affiliation{
Department physics, Faculty of science, Tokyo University of Science, \\
1-3 Kagurazaka, Shinjuku-ku, Tokyo 162-8601, Japan}
\date{\today}

\begin{abstract}
We study sound propagation in Bose-condensed gases in a highly-elongated harmonic trap at finite temperatures. This problem is studied within the framework of Zaremba-Nikuni-Griffin (ZNG) formalism, which consistent of a generalized Gross-Pitaevskii (GP) equation for the condensate and the kinetic equation for a thermal cloud. We extend the ZNG formalism to deal with a highly-anisotropic trap potential, and use it to simulate sound propagation using the trap parameters corresponding to the experiment on sound pulse propagation at finite temperature. We focus on the high-density two-fluid hydrodynamic regime, and explore the possibility of observing first and second sound pulse propagation. The results of numerical simulation are compared with an analyitical results derived from linearized ZNG hydrodynamic equations. We show that the second sound mode makes a dominant contribution to condensate motion in relatively high temperature, while the first sound mode makes an appreciable contribution. 
\end{abstract}

\pacs{}

\maketitle

\section{Introduction}
One of the phenomena attracting attention is the superfluid dynamics in ultracold atomic gases. 
Recently, there has been renewed interest in second sound mode in superfluid Bose and Fermi gases \cite{T_Heiselberg_Unitarysound,T_C,E_Joseph_PRL98,E_second_Bose,E_second_BF}. 
The existence of second sound is the most dramatic effects related to superfluidity in superfluid Bose and Fermi gases, which are described by Landau's two-fluid hydrodynamics analogous to the case of liquid $^4$He \cite{B_GNZ}. 
These equations describe the dynamics when collisions are sufficiently strong to produce a state of local thermodynamic equilibrium \cite{Hydro_Landau}. In this regime first and second sound modes
can be distinguished. The occurrence of two distinct modes is
caused by the presence of both superfluid component and normal
fluid component, which are coupled to each other. The study of ultracold gases in collisional hydrodynamic regime has been
difficult because the density and the $s$-wave scattering
length are typically not large enough. 
In the case of superfluid Fermi gases, Feshbach resonances allow ones to achieve conditions where the Landau two-fluid description is correct. 
Recent experiments have observed sound propagation in trapped superfluid Fermi gases with a Feshbach resonance \cite{E_Kinast_PRL92,E_Bartenstein_PRL92,E_Joseph_PRL98}.  

We note that the occurrence of two sound modes is caused by the coupled motion
of the superfluid component and normal fluid component. First sound is essentially
an in-phase oscillation of superfluid and normal fluid components, while second sound is 
an oscillation of two components. This is a general feature of first and second
sound, which is valid both for a dilute Bose gas and for superfluid $^4$He.
However, the detailed characteristics and behaviors of two sound modes are quite 
different in a Bose gas and in superfluid $^4$He.
In superfluid $^4$He, first sound is essentially a pressure wave, while second sound is essentially a temperature wave. In this case, second sound is completely uncoupled to the density fluctuations.
The situation is quite different in a dilute Bose gas, as discussed in Ref.~\cite{T_Griffin_PRA56}.  At very low temperature, the first sound mode is essentially the condensate collective mode and the second sound mode is the collective mode of quasiparticle excitations. 
With increasing temperature, hybridization two modes occurs,
and the nature of the sound oscillations changes.
At higher temperature, the first sound mostly involves the noncondensate oscillation, while the second sound mostly involves the condensate oscillation. In the case of a Bose gas, both first and second sound modes are coupled to density fluctuations. Therefore, the second sound pole makes a significant
pole to the dynamic density response function. This means that, in contrast to the case of
superfluid $^4$He, one can probe second sound in a dilute Bose gas by density perturbation.

Experimentally, sound wave in a highly-elongated trapped gas can be excited by a sudden modification of a trapping potential using the focused laser beam. 
The resulting density perturbations propagate with a speed of sound. 
This type of sound pulse experiment	was first carried out by MIT group for a Bose gas to probe Bogoliubov sound \cite{E_Andrews_PRL79}. 
Observed sound velocity was in
good agreement with theoretical predictions \cite{T_stringari_PRL}.  
In the case of trapped Fermi gases, first sound has been observed by the sound pulse propagation experiment \cite{ E_Joseph_PRL98}. 
Theoretically, sound pulse propagation in a trapped Bose gas have been studied for $T=0$ using Gross-Pitaevskii equation \cite {T_stringari_PRL,E_second_BF,T_Griffin_PRA56,T_Jacson_PRA75} and for a normal phase using the hydrodynamic equation \cite{T_Arahata_PRA2011}. 
The sound propagation was also studied theoretically for a normal Fermi gas using the kinetic equation \cite{T_tau_Nikuni}. The sound pulse propagation in superfluid Fermi gases in the two-fluid hydrodynamic regime was studied in Ref.~\cite{T_Ara_PRA2009_F}. In this regime, it was shown that two types of sound pulses, corresponding to first and second sound, propagate with their sound velocities.

More recently, sound propagation in Bose-condensed gases has been observed in Ref.~\cite{E_second_Bose} when the thermal cloud is in the hydrodynamic regime and the system is therefore described by the two-fluid model by using highly-elongated (cigar-shaped) traps. 
This experimental work reported evidence for a second sound mode in superfluid Bose gases, but first sound mode was not clearly identified. For completeness of the two-fluid hydrodynamics, it will be important to observe both first and second sound. 

In this paper, we study sound pulse propagation in Bose-condensed gases in a highly-elongated harmonic trap at finite temperatures.
In order to simulate the coupled motion of the condensate and noncondensate components in a fully consistent manner, we use the formalism developed by Zaremba, Nikuni, and Griffin (ZNG) \cite{T_ZNG_JLTP,B_GNZ}, that consists of a generalized Gross-Pitaevskii (GP) equation for a condensate and a kinetic equation for the thermal component. ZNG equations treat the excitations semiclassically within the Hartree-Fock (HF) approximation. Thus, the excitations dynamics with a thermal cloud of particles is governed by a Boltzmann equation for the phase-space distribution function. The coupled GP and Boltzmann equations include the transfer of atoms into and out of the condensate, which is taken
together with mean-field coupling between the two components. 

We will present several dynamical simulations of sound propagation based on the ZNG formalism. 
The procedure involves solving simultaneously a GP equation for the condensate and a Boltzmann kinetic equation for the thermal cloud. 
The sound pulse is excited by the same manner as the experiment way ~\cite{E_second_Bose}. In the case of a trapped Bose gas, one might think that the thermal density perturbations (first sound) is so small that one cannot distinguish small density perturbations from signal-to-noise in the thermal cloud. However, we will show that both first and second sound mode can be observed by a sudden modification of a trapping potential at intermediate temperatures. 

Since we are interested in the collision-dominated hydrodynamic regime, we have to simulate the system with a large number of thermal cloud atoms in order to achieve high enough density. However, numerical simulation of the ZNG equations for the system with a large number of thermal cloud atoms is very time consuming.   
In the present study, in order to save cost of numerical calculation, we derive quasi-1D ZNG equations by expanding the field operator in radial modes of the trap potential \cite{T_Ara_PRA2008}.
As shown in Ref.~\cite{T_Ara_PRA2008}, even when the dynamics of Bose-condensed gases in a highly-elongated harmonic trap is well approximated by 1-dimensional (1D) GP equation, the momentum space of the thermal cloud must be treated as three dimensional (3D), because the thermal cloud atoms typically have kinetic energy much larger than the typical energy associated with the radial trap frequency. Combining the work in Ref.~\cite{T_Ara_PRA2008} with the ZNG kinetic theory \cite{T_ZNG_JLTP,B_GNZ}, we develope the quasi-1D kinetic theory that include the degree of freedom in the radial direction. 

In Sec. II, we introduce quasi-1D ZNG equations appropriate for a highly-elongated Bose gas. 
The discussion closely follows the original approach given by Zaremba et al \cite{T_ZNG_JLTP,B_GNZ}. 
In this formalism, the condensate is described by a generalized quasi-1D GP equation for the Bose order parameter. It involves terms that are coupled to the noncondensate component. 
As in Refs~\cite{B_GNZ} and \cite{T_ZNG_JLTP}, we restrict ourselves to finite temperatures high enough that noncondensate atoms can be described by a semiclassical kinetic equation for the single-particle distribution function. 

In Sec. III, we show dynamical simulations for a Bose condensed gas in a highly-elongated harmonic trap with parameters corresponding to the experiment. We also estimate the collisional relaxation rate which defines the two-fluid hydrodynamic regime.

In Sec. IV, we discuss the first and second sound amplitude for condensate and noncondensate components separately using linearized ZNG hydrodynamic equations. In this section, we consider a uniform Bose condensed gas for simplicity. We calculate the relative weights of first and second sound mode using HF approximation for calculating thermodynamic various variables and compare those calculating by the dynamical simulation of the coupled ZNG equations.  
\section{Quasi 1D ZNG equations of a Bose Condensed Gas in a highly-elongated Harmonic Trap}
We consider a Bose condensed gas confined highly-elongated harmonic trap potential. Our system is described by the following
Hamiltonian :
\begin{eqnarray}
\hat H=\int d{\bf{r}}\left\{\hat {\psi}^\dagger({\bf r},t)\left[-\frac{\hbar^2}{2M}\nabla^2+V_{\rm ext}({\bf r})\right]\hat{\psi}({\bf r},t)+\frac{g}{2}\hat {\psi}^\dagger({\bf r},t)\hat {\psi}^\dagger({\bf r},t)\hat{\psi}({\bf r},t)\hat{\psi}({\bf r},t)\right\},\label{Eq_1}
\end{eqnarray}
with an anisotropic harmonic potential $V_{\rm ext}({\bf r})=\frac{M}{2} [\omega_{\bot}(x^2+y^2)+\omega_z z^2]$. In this paper, we consider a highly-elongated trap potential $\omega_z\ll \omega_{\bot}$. As usual, we treat the interatomic interaction in the $s-$wave approximation with $g=4\pi \hbar^2 a/M$, where $a$ is the $s-$wave scattering length and $M$ is an atomic mass.
In order to separate the radial and longitudinal degree of freedom, we expand the field operator in terms of the radial wavefunction\cite{T_Ara_PRA2008} 
\begin{eqnarray}
\hat \psi({\bf r},t)=\sum_{n}\hat \psi_n(z,t)\phi_n (x,y),\label{Eq_2}
\end{eqnarray}
where $\phi_n(x,y)$ is the normalized eigenfunction of the radial part of the single-particle Hamiltonian, which satisfies
\begin{eqnarray}
\left[-\frac{\hbar^2 }{2M}\nabla^2_{\bot}+\frac{M}{2}\omega_\bot (x^2+y^2)\right]\phi_n(x,y)=E_n\phi_n(x,y),\label{Eq3}
\end{eqnarray}
and $\hat{\psi}_n(z,t)$ satisfies the following equal time commutation relation :
\begin{eqnarray}
\left[\hat{\psi}_n(z,t),\hat{\psi}_{n^\prime}^\dagger(z^\prime,t)\right]=\delta(z-z^\prime)\delta_{n,n^\prime}.
\end{eqnarray}

Using (\ref{Eq_2}) and (\ref{Eq3}) in (\ref{Eq_1}), we rewrite the Hamiltonian as  
\begin{eqnarray}
\hat H&=&\sum_{n}\int dz \hat \psi^\dagger_n(z,t)\left[-\frac{\hbar^2}{2M}\frac{\partial^2}{\partial z^2 }+V_{ext}(z) +E_n\right]\hat \psi_n(z,t)\nonumber \\
&&+\sum_{nmlk}\frac{g_{nmlk}}{2}\int dz \hat  \psi^\dagger_n(z,t) \hat \psi^\dagger_m(z,t) \hat \psi_l(z,t) \hat \psi_k(z,t),\label{Eq_Quasi_2}
\end{eqnarray}
where the renormalized coupling constant is defined by 
\begin{eqnarray}
g_{nmlk}\equiv g\int dx \int dy \phi_n^\ast(x,y)\phi^\ast_m(x,y)\phi_l(x,y)\phi_k(x,y).
\end{eqnarray}
The Heisenberg equation of motion for the quantum field operator $\hat \psi_n(z,t)$ is given by
\begin{eqnarray}
i\hbar \frac{\partial }{\partial t}\hat \psi_n(z,t)&=&\left[\hat \psi_n(z,t),H\right]\nonumber \\
&=&\left[-\frac{\hbar^2}{2M}\frac{\partial^2}{\partial z^2 }+V_{\rm ext}(z) +E_n\right]\hat\psi_n(z,t)+\sum_{mlk}g_{nmlk} \hat \psi^\dagger_m(z,t) \hat \psi_l(z,t) \hat \psi_k(z,t). \label{1D_5}
\end{eqnarray}

In order to deal with the Bose broken symmetry, we separate out the condensate wavefunction from
the field operator as 
\begin{eqnarray}
\hat\psi_n(z,t)=\Phi_n(z,t)+\tilde \psi_n(z,t),
\end{eqnarray}
where the condensate wave function is defined by $\Phi_n(z,t)\equiv \langle \hat \Psi_n (z,t) \rangle $. 
The equation of motion for $\Phi_n$ can be obtained by taking statistical average of  (\ref{1D_5}) :
\begin{eqnarray}
i\hbar\frac{\partial }{\partial t}\Phi_n(z,t)&=&\left[-\frac{\hbar^2}{2M}\frac{\partial^z}{\partial z^2}+V_{\rm ext}(z)+E_{n}\right]\Phi_n(z,t)+\sum_{kl}\left(g_{nkkl}n_c^{k}+2g_{nkkl}\tilde n^{k}\right)\Phi_l(z,t)\nonumber \\
&&+\sum_{mk}g_{nkkl}\tilde m^{k}\Phi_l^{\ast}(z,t)+\sum_{mkl}g_{nmkl}\langle\tilde \psi_{m}^\dagger(z,t)\tilde \psi_{k}(z,t) \tilde \psi_{l}(z,t)\rangle ,
\label{1D_7}
\end{eqnarray}
where $n_c^{k}=\Phi^\ast_{k}(z)\Phi_{k}(z)$, $\tilde n^{k}=\langle\tilde \psi_{k}^\dagger(z,t) \tilde \psi_{k}(z,t) \rangle$.
In Eq.~(\ref{1D_7}) we have neglected the anomalous average $\langle\tilde \psi_{k}(z,t) \tilde \psi_{k}(z,t) \rangle$, as in Ref.~\cite{B_GNZ}. Moreover, we have assumed that the off diagonal terms of the noncondensate density, $\langle\tilde \psi_{k}^\dagger \tilde \psi_{k'} \rangle(k\neq k')$, are small and thus can be neglected. 
This assumption is expected to be valid in the case $\omega_z \ll \omega_\bot$ \cite{B_GNZ}. 
In addition, in the case where $\omega_z \ll \omega_\bot$, the contribution from higher radial modes to the condensate
wavefunction is negligibly small \cite{T_Ara_PRA2008}. Therefore, we will henceforth approximate $\Phi_{\alpha}=\Phi\delta_{\alpha, 0}$. 
With these approximations the generalized GP equation (\ref{1D_7}) reduces to 
\begin{eqnarray}
i\hbar\frac{\partial }{\partial t}\Phi(z,t)&=&\Bigg[-\frac{\hbar^2}{2M}\frac{\partial z}{\partial z^2}+V_{ext}(z)+E_{0}+g_{0000}n_c(z,t)
\nonumber \\&&~~~~~
+\sum_{k}2g_{0kk0}\tilde n^{k}(z,t)-iR(z,t)\Bigg]\Phi(z,t).
\label{GGP}
\end{eqnarray}
Here the source term $R$ is given by
\begin{eqnarray}
R(z,t)=\frac{\hbar\Gamma_{12}}{2n_c(z,t)},
\end{eqnarray}
with $\Gamma_{12}=-2{\rm Im} \left(\sum_{nmk}g_{0nmk}\Phi^\ast \langle\tilde \psi_{n}^\dagger(z,t)\tilde \psi_{m}(z,t) \tilde \psi_{k}(z,t)\rangle \right)$, $n_c(z,t)=|\Phi(z,t)|^2$ and $\tilde n^{k}=\langle\tilde \psi_{k}^\dagger(z,t) \tilde \psi_{k}(z,t) \rangle$. 

   We now turn to the dynamics of the noncondensate. The
physical properties of interest are in principle defined by the following
equation of motion obtained from (\ref{1D_5}) and (\ref{1D_7}):
\begin{eqnarray}
i\hbar \frac{\partial }{\partial t}\tilde \psi_{n} (z,t)&=&\left[-\frac{\hbar^2}{2M}+V_{\rm ext}(z)+E_n\right]\tilde \psi_{n} (z,t)+2\sum_{kl}g_{nkkl}n^{k}(z,t)\tilde \psi_l(z,t)\nonumber \\ &&
-2\sum_{kl}g_{nkkl}\tilde n^{k}(z,t)\tilde \psi_l(z,t)
+\sum_{l}g_{n00l}\Phi(z,t)\Phi(z,t)\tilde \psi_l(z,t)\nonumber \\ &&
+\sum_{mk}g_{nmk0}\Phi(z,t)\left[\tilde \psi_m(z,t) \tilde \psi_k(z,t)  -\tilde m^{k}(z,t)\delta_{mk}\right]\nonumber \\ &&
+2\sum_{mk}g_{nmk0}\Phi(z,t)\left[\tilde \psi^\dagger_m(z,t) \tilde \psi_k(z,t)  -\tilde n^{k}(z,t)\delta_{mk}\right]\nonumber \\ &&
+\sum_{mkl}g_{nmkl}\left[\tilde \psi^\dagger_m(z,t) \tilde \psi_k(z,t)\tilde \psi_l(z,t)-\langle\tilde \psi^\dagger_m(z,t) \tilde \psi_k(z,t)\tilde \psi_l(z,t)\rangle\right],
\end{eqnarray}
where $n^{k}=n_c\delta_{0,k}+\tilde n^{k}$. It is convenient to define the time evolution of $\tilde \psi_n(z,t)$ by
\begin{eqnarray}
\tilde \psi_n(z,t)=S^{\dagger}(t,t_0) \tilde \psi_n(z,t) S(t,t_0), \label{1D_11}
\end{eqnarray}
where the unitary operator $S(t,t_0)$ evolves according to the equation of
motion
\begin{eqnarray}
i\hbar \frac{d}{d t}S(t,t_0)=\hat H_{\rm eff}S(t,t_0)  ,\label{1D_11}
\end{eqnarray}
with $S(t_0,t_0)$=1.
The effective Hamiltonian in (\ref{1D_11}) is given
by 
\begin{eqnarray}
\hat H_{\rm eff}&=&\hat H_0+\hat H^\prime,\nonumber \\
\hat H^\prime&=& \hat H_1^\prime+\hat H_2^\prime+\hat H_3^\prime+\hat H_4^\prime,
\end{eqnarray}
where the various contributions are defined as
\begin{eqnarray}
\hat H_0&=&\sum_n \int dz \tilde \psi_n^\dagger (z,t)\left(-\frac{\hbar^2 }{2M}\frac{\partial^2 }{\partial z^2}+E_n +\tilde \psi_n^\dagger U^{n}(z,t) \right)\tilde \psi_n (z,t), \\
U^{n}(z,t)&=&V_{\rm ext}(z)+2\sum_{k}g_{nkkn}n^{k}(z,t),\\
\hat H_1^\prime&=&\sum_{n}\int dz \left(L_1\tilde \psi_n^\dagger+L_1^* \tilde \psi_n \right),\\
L_1&=&\sum_{mkl}g_{nmkl}\left[2\tilde n^{k}\Phi\delta_{0,l}\delta_{mk}+\tilde{m}^{k}\Phi^\ast\delta_{0,l}\delta_{mk}+\langle \tilde \psi_{m}^\dagger\tilde \psi_{k}\tilde\psi_{l}\rangle\right],\\
\hat H_2^\prime&=&\sum_{kl}\frac{g_{00kl}}{2}\int dz \left(\Phi\Phi\tilde \psi_k^\dagger \tilde \psi_l^\dagger+\Phi^\ast\Phi^\ast\tilde \psi_k \tilde \psi_l\right),\\
\hat H_3^\prime&=&\sum_{mkl}g_{0mk0}\int dz \left(\Phi\tilde \psi_m^\dagger\tilde \psi_k^\dagger \tilde \psi_l+\Phi^\ast \tilde \psi_m^\dagger \tilde \psi_k \tilde \psi_l\right), \\
\hat H_4^\prime&=&\sum_{nmkl}\int dz \left(\frac{g_{nmkl}}{2}\tilde \psi_n^\dagger\tilde \psi_m^\dagger\tilde \psi_k \tilde \psi_l-2g_{nnkl}\tilde n^{n} \tilde \psi_k^\dagger \tilde \psi_l\right).
\end{eqnarray}
The expectation value of an ordinary operator defined in terms of $\hat \psi_n$ and $\hat \psi_n^\dagger$ is given by  
\begin{eqnarray}
\langle \hat O(t) \rangle\equiv  \langle \hat O \rangle_t={\rm Tr} \hat \rho_0(t_0)\hat O(t)={\rm Tr} \tilde \rho (t,t_0)\hat O(t_0),
\end{eqnarray}
where $\tilde \rho (t,t_0)=\hat S^\dagger (t,t_0)\hat \rho(t_0)\hat S (t,t_0)$
satisfies the following equation
\begin{eqnarray}
i\hbar\frac{d\tilde\rho(t,t_0)}{dt}=\left[\hat H_{\rm eff},\tilde\rho(t,t_0)\right].
\label{ZNG16}
\end{eqnarray}

Our ultimate objective is to obtain a quantum kinetic equation for the
noncondensate atoms. 
We define the Wigner operator as
\begin{eqnarray}
\hat f_{n}(p_z,z,t_0)&=&\int dz^\prime  e^{i{p_z}\cdot {z}^\prime/\hbar}
 \tilde{\psi}^\dagger_n\left({z}+\frac{z^\prime}{2},t_0\right)\tilde{\psi}_n\left(z-\frac{z^\prime}{2},t_0\right).
\end{eqnarray}
The Wigner distribution function is then given by
\begin{eqnarray}
f_{n}(p_z,z,t)={\rm Tr} \tilde{\rho}(t,t_0)\hat{f}_{n}(p_z,z,t_0).
\end{eqnarray}
The equation of motion for $f$ is obtained by using Eq.~(\ref{ZNG16})
\begin{eqnarray}
\frac{\partial f_{n}(p_z, z,t)}{\partial t}&=&\frac{1}{i\hbar}{\rm Tr}\tilde\rho(t,t_0)[\hat f_{n}(p_z, z,t_0),H_{0}(t)]+\frac{1}{i\hbar}{\rm Tr}\tilde\rho(t,t_0)[\hat f_{n}(p_z, z,t_0),H^\prime(t)].
\end{eqnarray}
With the assumption that $U^{n}(z, t)$ varies slowly in space, we then have 
\begin{eqnarray}
[\hat f_{n}(p_z, z,t_0),H_{0}(t)]\simeq-\frac{i\hbar}{M}p_z\frac{\partial }{\partial z} \hat f_{n}(p_z, z,t_0)+i\hbar \frac{\partial }{\partial z}U^{n}(z,t)\frac{\partial }{\partial p_z}\hat{f}_{n}(p_z, z,t_0).
\label{eq_f28}
\end{eqnarray}
The second right hand side of the right hand side of Eq.~(\ref{eq_f28}) represents the effect of collisions
between the atoms. As we show in Appendix A, the collision
integral is the sum of two contributions:
\begin{eqnarray}
\frac{\partial f_{n}}{\partial t}\bigg|_{\rm coll}=C_{12}[f_{n}]+C_{22}[f_{n}].
\end{eqnarray}
Thus we obtain
\begin{eqnarray}
\frac{\partial f_{n}(p_z, z,t)}{\partial t}+\frac{p_z}{M}\frac{\partial }{\partial z} f_{n}(p_z, z,t)- \frac{\partial }{\partial z}U^{n}(z,t)\frac{\partial }{\partial p_z}{f}_{n}(p_z, z,t)\nonumber \\
=C_{12}[f_{n}]+C_{22}[f_{n}].\label{Bolt}
\end{eqnarray}
The $C_{12}$  collision integral is defined as the contribution from the $H^\prime_3$ perturbation
\begin{eqnarray}
C_{12}[f_{n}]&\equiv&-i{\rm Tr} \rho(t,t_0)\left[\hat{f}_{n}(p_z,z,t_0),\hat H_{3}^\prime(t)\right]\nonumber\\ 
&=&4\pi \sum_{n'm'k'}g_{n'm'k'0}^2n_c 
\sum_{p_{z1},p_{z2},p_{z3}}\left[\delta(\epsilon_c+\tilde{\epsilon}_1^{n'}-\tilde{\epsilon}_2^{m'}-\tilde{\epsilon}_3^{k'})\right]\delta_{p_{zc}+p_{z1},p_{z2}+p_{z3}}
\nonumber \\&&
(\delta_{p_z,p_{z1}}\delta_{nn'}-\delta_{p_{z},p_{z2}}\delta_{nm'}-\delta_{p_{z},p_{z3}}\delta_{nk'})[(1+f_1^{n'})f_2^{m'}f_3^{k'}-f_1^{n'}(1+f_2^{m'})(1+f_3^{k'})], 
\nonumber  \\ \label{C_12}
\end{eqnarray}
where the local HF single-particle energie is  
$\tilde{\epsilon}^l=\frac{p_{z}^2}{2M}+E_l+U^{l}(z,t)$ and $\epsilon_c=\mu_c+\frac{1}{2}mv_c^2$.
The local condensate chemical potential $\mu_c$ is defined by
\begin{eqnarray}
\mu_c&\equiv&-\frac{\hbar^2}{2M\sqrt{n_c(z,t)}}\frac{\partial }
{\partial z}\sqrt{n_c(z,t)}+V_{\rm ext}+E_0+g_{0000}n_c+\sum_{k}2g_{0kk0}\tilde n^{k},
\end{eqnarray}
and the condensate velocity is given by $v_c\equiv \frac{\hbar}{M}\frac{\partial }
{\partial z}\theta(z,t)$ with $\Phi(z,t)=\sqrt{n_c(z,t)}e^{i\theta(z,t)}$.
The source term $R$ is directly related to the $C_{12}$ collision term 
\begin{eqnarray}
R(z,t)=\frac{\hbar}{2n_c(z,t)}\sum_n\int \frac{dp_z}{2\pi \hbar} C_{12}[f_n].
\end{eqnarray}
Similarly, the $C_{22}$ collision is defined as the $H_4'$ perturbation, which is obtained as
\begin{eqnarray}
C_{22}[f_{n}]&\equiv&-i{\rm Tr} \rho(t,t_0)\left[\hat{f}_{n}(p_z,z,t_0),\hat H_{4}^\prime(t)\right]\nonumber\\ 
&=&\pi \sum_{n'm'k'l'}g_{n'm'k'l'}^2
\sum_{p_{z1},p_{z2},p_{z3},p_{z4}}\left[\delta(\tilde{\epsilon}_1^{n'}-\tilde{\epsilon}_2^{k'}-\tilde{\epsilon}_3^{m'}-\tilde{\epsilon}_4^{l'})\right]
\nonumber \\&&
\delta_{p_{z1}+p_{z2},p_{z3}+p_{z4}}
(\delta_{p_z,p_{z1}}\delta_{nn'}+\delta_{p_{z},p_{z2}}\delta_{nm'}-\delta_{p_{z},p_{z3}}\delta_{nk'}-\delta_{p_{z},p_{z4}}\delta_{nl'})\nonumber \\&&
[(1+f_1^{n'})(1+f_2^{k'})f_3^{m'}f_4^{l'}-f_1^{n'}f_2^{k'}(1+f_3^{m'})(1+f_4^{l'})].
\label{C_22}
\end{eqnarray}
We refer to Appendix A for detail derivations of the collision integrals $C_{12}$ and $C_{22}$.

In summery, we have obtained a coupled set of equation of motion for the condensate and noncondensate as follows:
\begin{eqnarray}
&&i\hbar\frac{\partial }{\partial t}\Phi(z,t)=\Bigg[-\frac{\hbar^2}{2M}\frac{\partial z}{\partial z^2}+V_{ext}(z)+E_{0}+g_{0000}n_c(z,t)
\nonumber \\&&~~~~~~~~~~~~~~~~
+\sum_{k}2g_{0kk0}\tilde n^{k}(z,t)-iR(z,t)\Bigg]\Phi(z,t), \label{sumGP}
\\ 
&&\frac{\partial f_{n}(p_z, z,t)}{\partial t}+\frac{p_z}{M}\frac{\partial }{\partial z} f_{n}(p_z, z,t)- \frac{\partial }{\partial z}U^{n}(z,t)\frac{\partial }{\partial p_z}{f}_{n}(p_z, z,t)\nonumber \\ &&~~~~~~~~~~~~~~~~~~~~~~~~~~~~~~~~~~~~~~~~~~~~~~~~~~
=C_{12}[f_{n}]+C_{22}[f_{n}].
\end{eqnarray}
The condensate is described by a quasi-1D GP equation for $\Phi(z,t)$. The noncondensate is described by a quasi-1D kinetic equation for the distribution function $f^n(p_z,z,t)$. Here $n$ is the radial mode index. Different radial mode are coupled through the mean-field interaction as well as collisions.   

Before closing section, we give equilibrium solution of the coupled ZNG equations.   
The equilibrium solution for the condensate wavefunction is given by $\Phi_0(z,t)=\Phi_0(z)e^{-i\mu_{c0}t/\hbar}$, where $\Phi_0(z)$ satisfies
 \begin{eqnarray}
\Bigg[-\frac{\hbar^2}{2M}\frac{\partial z}{\partial^2}+V_{ext}(z)+E_{0}+g_{0000}n_{c0}(z)
+\sum_{k}2g_{0kk0}\tilde n_0^{k}(z,t)\Bigg]\Phi_0(z)
=\mu_{c0}\Phi_0(z).\label{S_GP}
\end{eqnarray}
Here $\mu_0$ is equilibrium chemical potential. 
The equilibrium distribution function is given by the static equilibrium Bose distribution \begin{eqnarray}
f_n^0(p_z, z)=\frac{1}{\exp \{\beta_0[p^2/2M+U^n_0(z)-\mu_{c0}]\}-1},
\label{S_f}
\end{eqnarray}
where $\beta_0=\frac{1}{k_{B}T_0}$ is the inverse uniform temperature. The trapping potential is augmented by the HF mean-field 
$U^n_0=V_{\rm ext}(z)+2\sum_{k}g_{nkkn}n_0^{k}(z)$. The coupled equations (\ref{S_GP}) and (\ref{S_f}) must be solved self-consistently. 

\section{Dynamics of First and Second Sound in a Bose Condensed gas}
Using the quasi-1D ZNG equations derived in the previous section, we study  sound pulse propagation excited by a sudden modification of a trapping potential.
The numerical procedure for calculating ZNG equations closely follows that described in Ref.~\cite{T_Jacson_Zaremba,B_GNZ}. 
 The dynamics of the thermal cloud is calculated by using $N$-body simulations \cite{T_Jacson_Zaremba}. 
The dynamics of the condensate is determined by numerically propagating the GP equation using a split-operator fast Fourier transform (FFT) method. 
The numerical method is described in detail in Appendix B. 

We take the physical parameters from the experiment of Ref.~\cite{E_second_Bose}, which reports the observation of second sound propagation. In this experiment, total number of $^{23}$Na atoms $N=1.7\times 10^8$, radial trap frequency $\omega_{\rm rad}/2\pi=95$Hz, and the aspect ratio $\omega_{\rm rad}/\omega_{\rm ax}\approx 65$. 
In this situation, one has a high density cloud of $n_0\sim10^{20}$ cm$^{-3}$. 
At the lowest temperatures, the BEC has a radial TF radius of roughly 22 $\mu$m and an axial TF radius of 1.4 mm. 
The number of test particles is ten times the actual number of thermal atoms in order to minimize the effects of a discrete particle description.

\begin{figure}[htbp]
\includegraphics[height=2.0in]{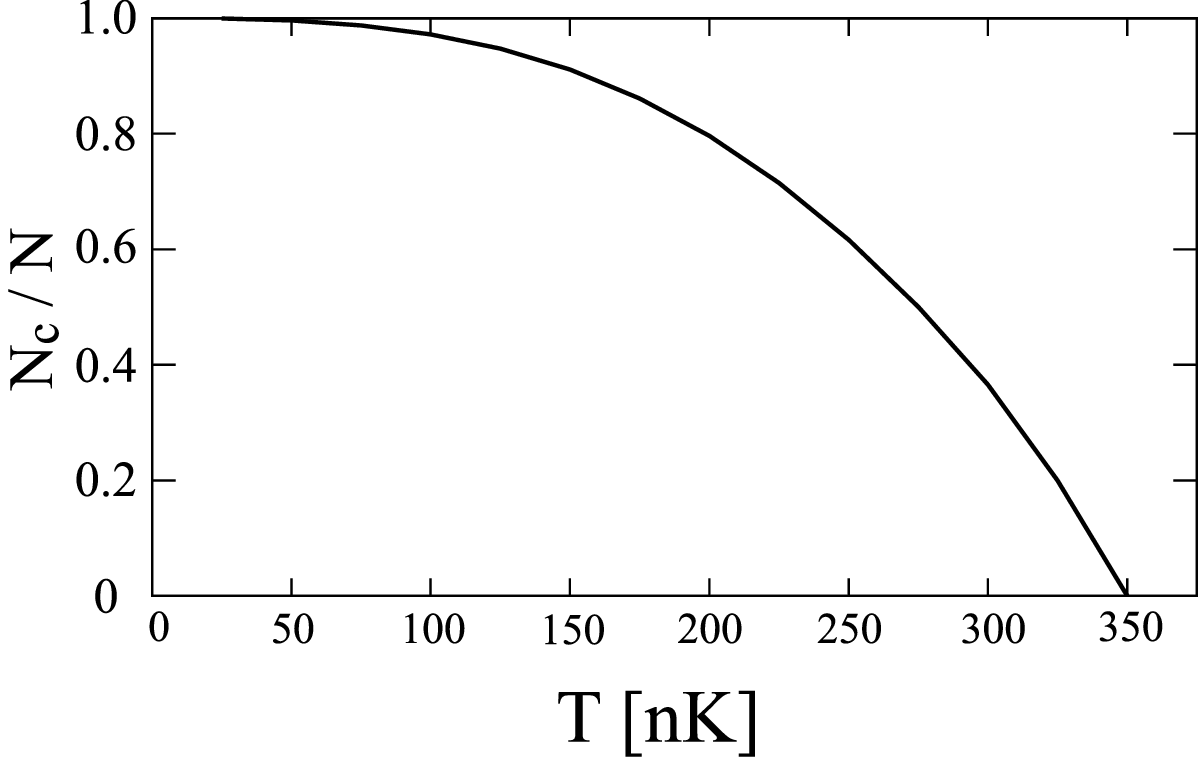}
 \caption{ The condensate fraction $N_{\rm{c}}/$$N$ as a function of the temperature.} 
  \label{Fig:Nc}
\end{figure}

\begin{figure}[htbp]
\includegraphics[height=2.5in]{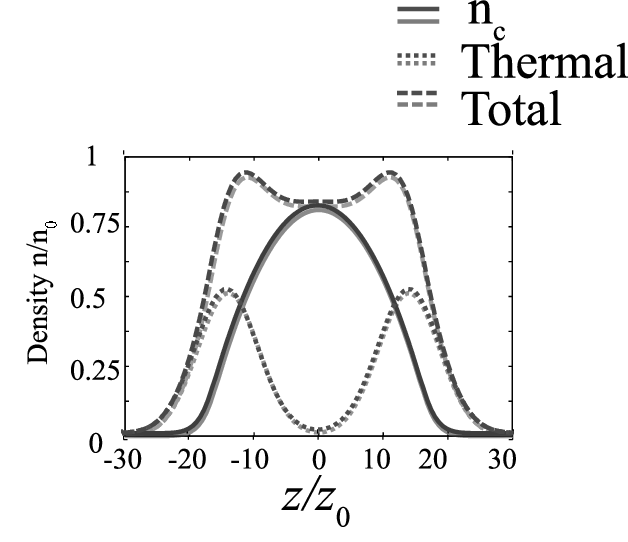}
 \caption{Axial density profiles in the equilibrium state. Gray lines are solutions within the quasi-1D approximation. Black line are equilibrium solutions without making the quasi-1D approximation.} 
  \label{Fig:hi}
\end{figure}

We first consider equilibrium solutions (\ref{S_GP}) and (\ref{S_f}) for these experimental parameters.
In Fig.~\ref{Fig:Nc}, we plot the condensate fraction $N_{\rm{c}}/N$ as a function of the temperature.
We see that the transition temperature for the Bose-Einstain condensation is given by $T_c\simeq 350 $nK.    
In Fig.~\ref{Fig:hi}, we plot the equilibrium density profiles of the condensate and noncondensate at $T=176.5$ nK($\simeq0.5 T_c$). 
For comparison, we also show equilibrium density profiles obtained from the full-3D HF calculation, i.e. without making the quasi-1D approximation, in Fig.~\ref{Fig:hi}. The differences between with and without the quasi-1D approximation are only a few \%.  
This confirms that our quasi-1D treatment can well describe the highly-elongated system. We note that in order to obtain reasonable results, we must take large enough number of radial modes so that $E_n > k_{\rm B} T$. For example, we took about 1000 radial mode for the calculate of Fig.~\ref{Fig:hi}.  

\begin{figure}[htbp]
\includegraphics[height=2.0in]{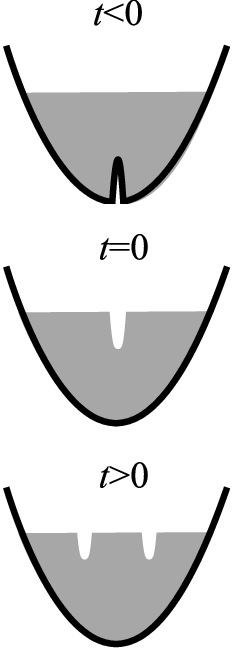}
 \caption{Schematic representation of the excitation of a sound wave, where the trapping potential, height and width of the perturbation are roughly on scale.} 
  \label{Fig:SO}
\end{figure}

\begin{figure}[htbp]
\includegraphics[height=3.0in]{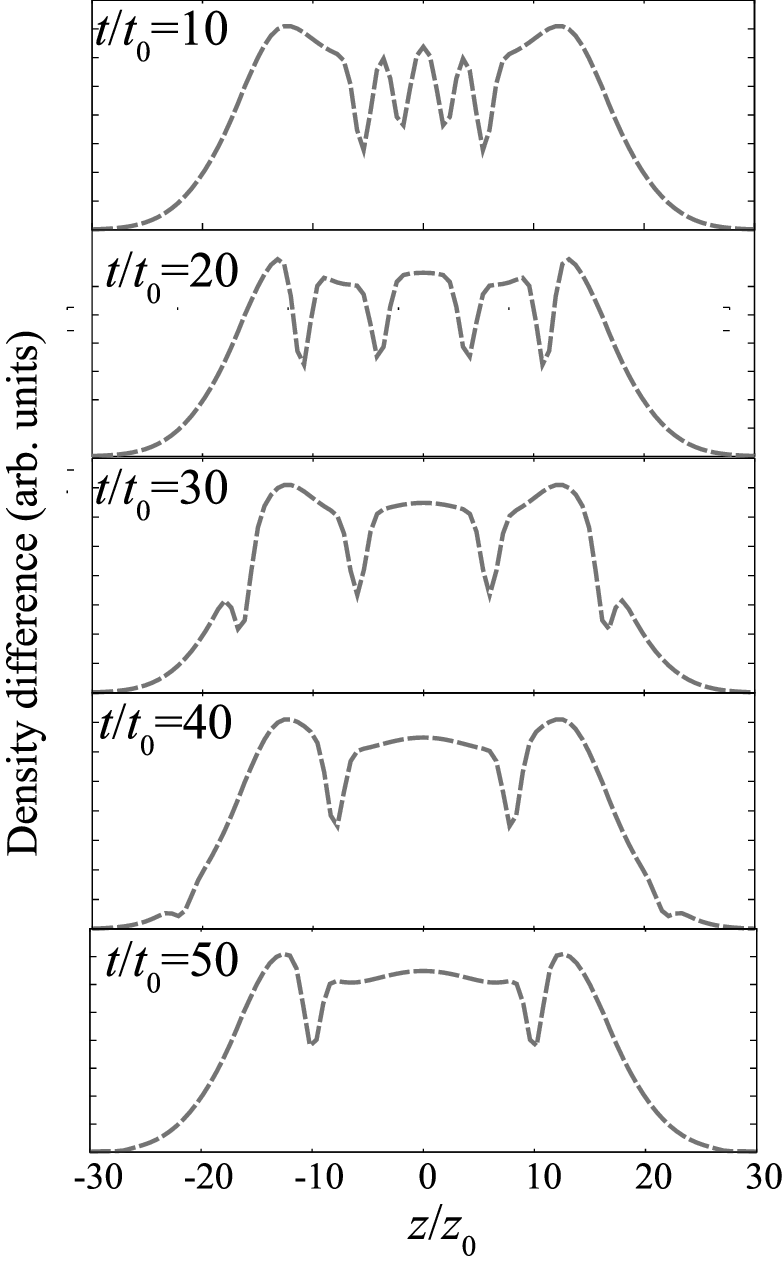}
 \caption{Axial density profiles of condensate and noncondensate for various propagation times at $T\simeq0.5T_{\rm c}$.} 
  \label{Fig_D_T=05}
\end{figure}

\begin{figure}[htbp]
\includegraphics[height=3.0in]{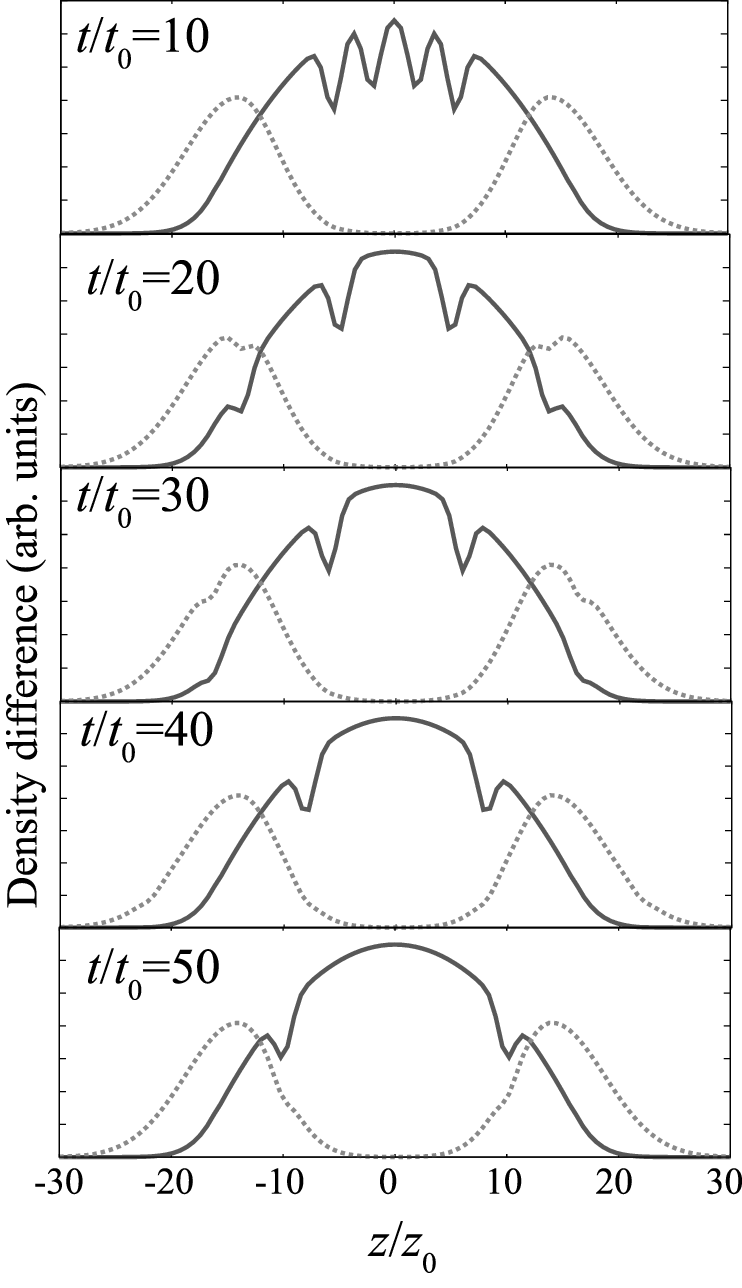}
 \caption{Axial density profiles of total density for various propagation times at $T\simeq0.5T_{\rm c}$.} 
  \label{Fig_D_S=05}
\end{figure}

We now consider density disturbance by a sudden modification of the external potential generating pulse propagation. Here we set the external potential $\delta U (z,t)=Ae^{-az^2} \theta(-t) $ with $A\sim 0.5 \mu_0 $ and $a\sim 250 \sqrt{z_0}$ with $z_0=\sqrt{M\omega /\hbar }$. A localized potential is applied at $t<0$, while it is turned off at $t=0$. This situation can be described as presence of a localized potential aimed at the center of the trap, which acts as a repulsive trap. Turning the potential suddenly off causes a local dip of the BEC density. This perturbation splits up in two waves propagating symmetrically outward, both with half the amplitude of the initial perturbation. A schematic representation of the excitation procedure is shown in Fig. \ref{Fig:SO}. 
The axial density profiles, shown in Fig. \ref{Fig_D_T=05} for various propagation times, clearly shows that density dips corresponding to two sound modes travel with their sound velocity. The faster sound pulse corresponds to the first sound, while the slower sound pulse corresponding to the second sound. We see that the both depth of the first and second sound dips at $T\simeq 0.5T_{\rm c}$ are sufficiently large for the experimental observation. 
For comparison, we plot axial density profiles of condensate and noncondensate separately  for various propagation times in Fig. \ref{Fig_D_S=05}. 

  \begin{figure}[htbp]
\includegraphics[height=2.0in]{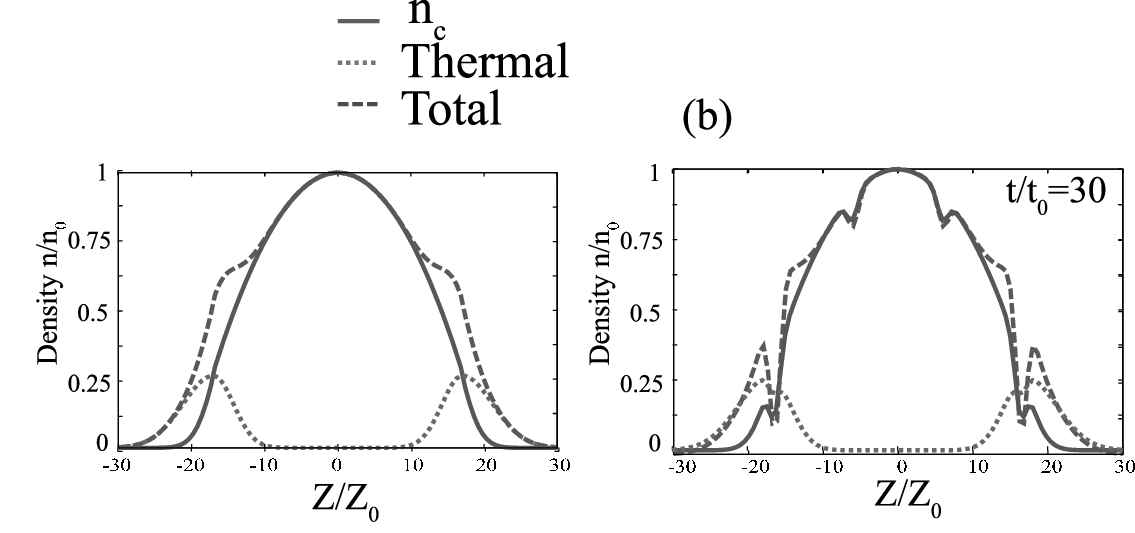}
 \caption{(a) Axial density profiles in the equilibrium state without perturbation (i. e., $\delta U=0$) at $T= 88.2$ nK($\simeq0.25 T_c$). (b) Axial density profiles at propagation time $t/t_0=30$. } 
  \label{Fig_D_T=025}
\end{figure} 
  \begin{figure}[htbp]
\includegraphics[height=2.0in]{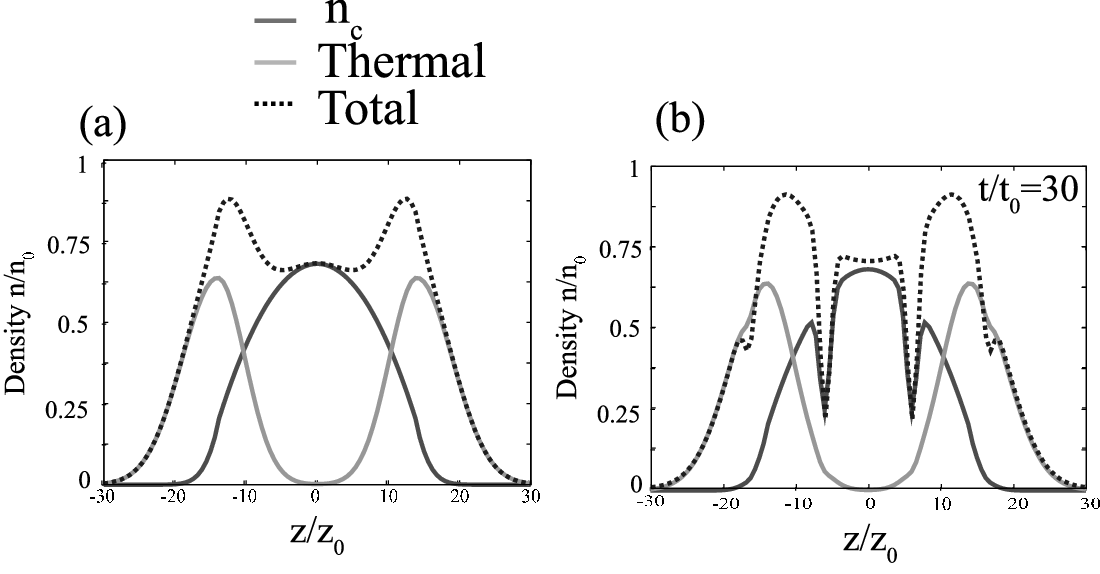}
 \caption{(a) Axial density profiles in the equilibrium state without perturbation (i. e., $\delta U=0$) at $T=264.5$ nK($\simeq0.75T_c$). (b) Axial density profiles at propagation time $t/t_0=30$} 
  \label{Fig_D_T=075}
\end{figure} 
The axial density profiles at $T= 88.2$ nK($\simeq0.25 T_c$) and $T=264.5$ nK($\simeq0.75T_c$) are shown in Figs. \ref{Fig_D_T=025} and \ref{Fig_D_T=075}. Compared to Fig.~\ref{Fig_D_S=05} with Fig.~\ref{Fig_D_T=025} and Fig.~\ref{Fig_D_T=075}, we see that the first sound mode is dominant at low temperature, while the second sound is dominant at high temperature. 

We note that it is difficult to observe sound propagation in a noncondensate thermal cloud because the thermal density perturbations is so small that one cannot distinguish small density perturbations from signal-to-noise in the thermal cloud. Nevertheless, at the intermediate temperature $T\simeq0.5 T_{\rm c}$ both first and second sound pulses appear in the {\it total} density. 
In this regard, we note that the analysis of Ref.~\cite{E_second_Bose} was based on the assumption that the condensate motion is always dominated by second sound at all temperatures. However, the calculation shows that the condensate motion is dominated by first sound at low temperatures. Therefore, it is possible that the experimental result of Ref.~\cite{E_second_Bose} at $T<0.5 T_{\rm c}$ may have observed first sound. It may therefore require a careful analysis of the experimental data in the crossover temperature regime $T\sim0.5 T_{\rm c}$ in order to identify two sound modes. 

\begin{figure}[htbp]
\includegraphics[height=2.0in]{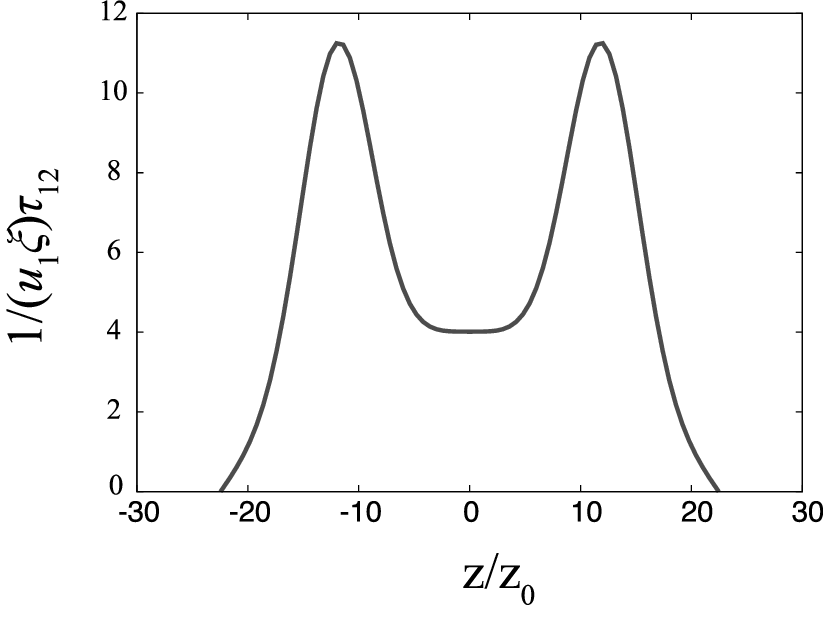}
 \caption{The equilibrium local collision rate $1/\tau_{12}$ in a trap (in units of a frequency $\omega_0\equiv u_1 \xi$ with the speed of first sound $u_1$ and healing length $\xi \equiv \frac{\hbar}{\sqrt{2mgn_c}}$) as a function of the $z$ distance (in units of the harmonic oscillator length) at $T
 \simeq0.75T_{\rm c}$.} 
  \label{Fig_tau}
\end{figure}

Let us now examine validity of two-fluid hydrodynamics in our system.
The existence of first and second sound is predicted by Landau two-fluid hydrodynamics, which is valid when collisions are sufficiently strong to produce a state of local thermodynamic equilibrium \cite{Hydro_Landau}. This requirement is usually summarized as $\omega \tau \ll 1$, where $\omega$ is the frequency of a collective mode and $\tau$ is the appropriate relaxation rate. 
In a trapped Bose gas, a relevant relaxation time is $\tau_{12}$ relaxation time associated with the $C_{12}$ collisions \cite{T_ZNG_JLTP, B_GNZ}, which describes equilibration between the condensate and the thermal cloud. The relation time $\tau_{12}$ is given by 
\begin{eqnarray}
1/\tau_{12}=\sum_n\frac{\Gamma_{12}^{\rm out}[f_n]}{\tilde n^n},
\end{eqnarray}
where 
\begin{eqnarray}
\Gamma_{12}^{\rm out}[f_n]&=&\sum_{n'm'k'}\frac{g_{n'm'k'0}^2n_c}{\pi\hbar^3}
 \int dp_{z2} \int dp_{z3}\int dp_{z4} \delta(\epsilon_c+\tilde{\epsilon}_2^{n'}-\tilde{\epsilon}_3^{m'}-\tilde{\epsilon}_4^{k'})\nonumber \\ && \times \delta({p_{cz}+p_{z2}-p_{z3}-p_{z4}})
(\delta_{nn'}-\delta_{nm'}-\delta_{nk'})f_2^{n'}(1+f_3^{m'})(1+f_4^{k'}).
\end{eqnarray}
In Fig.~\ref{Fig_tau}, we plot the equilibrium local collision rate $1/\tau_{12}$ in a trap as a function of the $z$ distance at $T\simeq 0.75T_{\rm c}$.
This collision rate again has a maximum at the edge if the condensate, falls off rapidly beyond this point, being proportional to the condensate density $n_c$. 
This figure shows that $\omega_0 \tau_{12}<1$ in the whole region of the condensate, 
where $\omega_0\equiv u_1\xi$, $\xi\equiv \frac{\hbar}{\sqrt{2mgn_c}}$ being the healing length. 
For the trap parameters given above and in the temperature range $0.25T_{\rm c}\lesssim T\lesssim0.7T_{\rm c}$,  we found that the equilibrium local collision rate satisfies $\omega_0 \tau_{12}<1$  in the whole region of the condensate. Thus, the sound propagation experiment is well within the hydrodynamic regime \cite{T_Arahata_PRA2011}.
For comparison, we also calculated $1/\tau_{12}$ without making the quasi-1D approximation $\frac{1}{\bar \tau_{12}(z)}=\int dx dy \frac{ \tilde n_0}{\tau_{12}^{3D}}/\int dx dy \tilde n_0$ \cite{B_GNZ}. The differences between with and without the quasi-1D approximation are only a few \%.  
This also confirm that our quasi-1D ZNG equation describe
dynamics of the highly-elongated system quite well.

\begin{figure}[htbp]
\includegraphics[height=3.in]{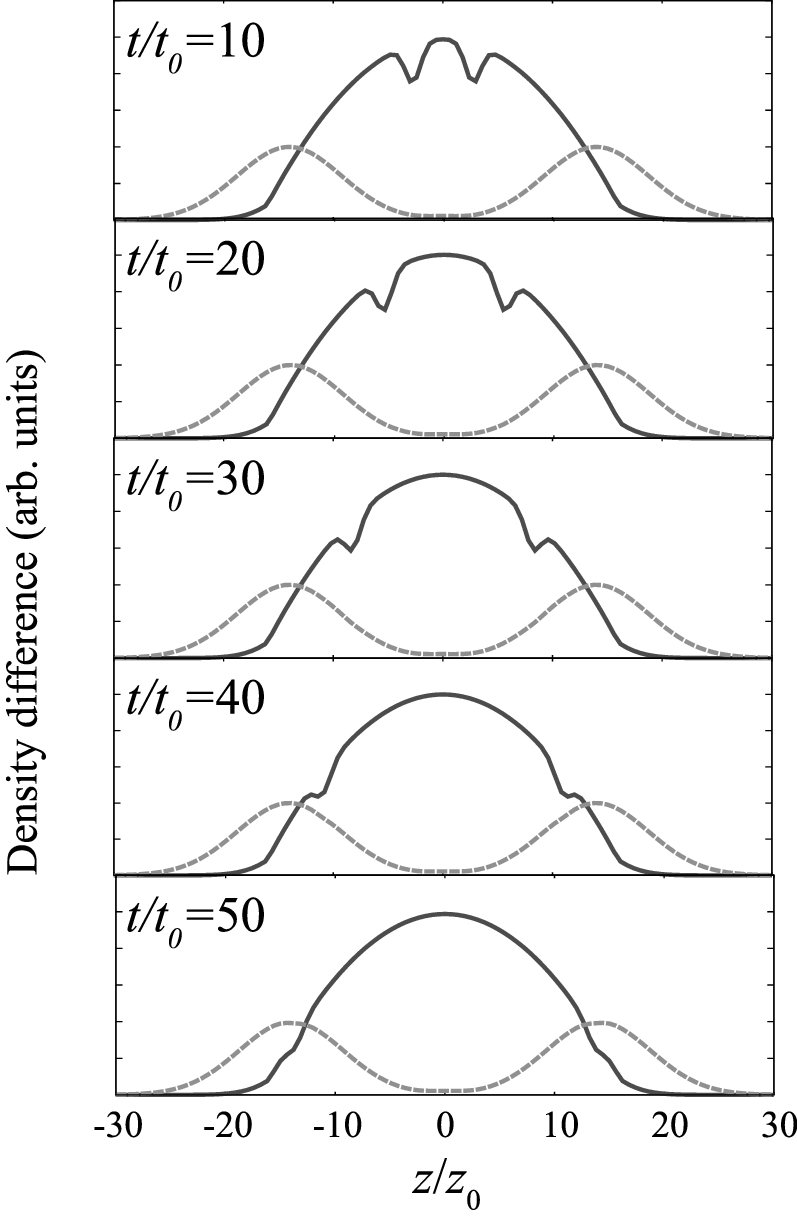}
 \caption{Axial density profiles of condensate and noncondensate for various propagation times for $N=10^5$ at $T\simeq 0.5T_{\rm c}$. In this case, the system is in the collisionless regime.} 
  \label{Fig:nonhydro}
\end{figure}

In the collisionless regime, i.e. $\omega_0 \tau_{12}>1$, the first  and second sound cannot be excited, but only Bogoliubov sound can be excited by a sudden modification of a
trapping potential. In Fig. \ref{Fig:nonhydro} we show the sound pulse propagation at total number of atoms $N=10^5$ corresponding to the density $n=10^{14}$ cm$^{-3}$ at $T\simeq 0.5 T_c$.
We see that the dynamics of the cloud is quite different from Fig.~\ref{Fig_D_S=05}. Clearly, only the Bogoliubov sound propagates. This situation is similar to the case of MIT experiment in Ref.~\cite{E_Andrews_PRL79}.

\section{Comparison with linear response solution of ZNG hydrodynamic equations for a uniform gas}
In this section, we derive an analytical expression for the amplitude of first and second sound pulses for a uniform gas. We limit ourself to the hydrodynamic regime. General expressions for pulse amplitudes have been derived using Landau two-fluid hydrodynamic equations in Ref.~\cite{T_Ara_PRA2009_F}. Here we instead use ZNG hydrodynamic equations, which allows for direct comparison with the simulation results.
We start with the linearized ZNG hydrodynamic equations for a
{\it uniform} gas \cite{T_ZNG_sound}:
\begin{eqnarray}
{\partial\delta \tilde{n}\over\partial t} &=& -
\tilde{n}_0({\nabla}\cdot\delta{\bf v}_n)+\delta \Gamma_{12}, \label{greens_eq1}\\
M\tilde{n}_0{\partial\delta{\bf v}_n\over\partial t} &=& -{\nabla}
\delta\tilde{P}-2g\tilde{n}_0{\nabla} (\delta\tilde{n}+\delta n_c)-\tilde{n}_0\nabla\delta U, 
\label{greens_eq2}\\
{\partial\delta\tilde{P}\over\partial t} &=& -{5 \over 3}
\tilde{P}_0 ({\nabla}\cdot\delta{\bf v}_n)
+{2\over 3} (\mu_{c0} - U_0) \delta
\Gamma_{12}\, ,
\label{greens_eq3}
\end{eqnarray}
\begin{eqnarray}
{\partial\delta n_c\over\partial t} &=& - 
n_{c0}  ( {\nabla}\cdot \delta{\bf v}_c)
-\delta \Gamma_{12}, \label{greens_eq4}\\
M{\partial\delta{\bf v}_c\over\partial t}
&=& -{\nabla}(\delta\mu_c-\delta U),
\label{greens_eq5}
\end{eqnarray}
\noindent
where
\begin{equation}
\delta \mu_c =  g\delta n_c
+ 2g\delta \tilde n \,.
\label{greens_eq6}
\end{equation}
The expression for $\delta \Gamma_{12}$ is given by
\begin{equation}
\delta \Gamma_{12}[\tilde f]=
-{\beta_0 n_{c0} \over \tau_{12}} \delta\mu_{\rm diff},~~~
\mu_{\rm diff} \equiv \tilde \mu - \mu_c,
\label{greens_eq7}
\end{equation}
\begin{eqnarray}
\mu_{c0}=gn_{c0}+2g\tilde{n}, \ U_0=2g(n_{c0}+\tilde{n}).
\end{eqnarray}
In the above equations, we have explicitly included the time-dependent external perturbation $\delta U({\bf r},t)$.
To solve the linearized hydrodynamic equations,
we introduce velocity potentials according to $\delta
{\bf v}_c \equiv {\nabla} \phi_c$ and $\delta {\bf v}_n \equiv {\nabla} \phi_n$.  In terms of these new variables, 
the equations for the condensate and the equations for the
noncondensate  can be combined to give
\begin{eqnarray}
M{\partial^2 \phi_c \over \partial t^2} &=& gn_{c0} \nabla^2 \phi_c +
2g\tilde n_0 \nabla^2 \phi_n +{\sigma_H\over \tau_\mu} \delta \mu_{\rm diff}-\frac{\partial }{\partial t}\delta U,
\label{greens_eq9} \\
M{\partial^2 \phi_n \over \partial t^2} &=& \left ( {5\tilde P_0 \over
3\tilde n_0} + 2g\tilde n_0 \right ) \nabla^2 \phi_n +
2g n_{c0} \nabla^2 \phi_c - {2\sigma_H \over 3
\tau_\mu} \frac{n_{c0}}{\tilde n_0}
\delta \mu_{\rm diff}-\frac{\partial }{\partial t}\delta U.
\label{greens_eq10}
\end{eqnarray}
Here $\delta\Gamma_{12}$ has been expressed in terms of  $\delta\mu_{\rm diff}$  using (\ref{greens_eq7}).
The equation of motion for $\delta\mu_{\rm diff}$ is given by
\begin{equation}
\frac{\partial \delta\mu_{\rm diff}}{\partial t}=
\frac{2}{3}gn_{c0}\nabla^2\phi_n-
gn_{c0}\nabla^2\phi_c
-\frac{\delta \mu_{\rm diff}}{\tau_{\mu}},
\label{greens_eq11}
\end{equation}
where the relation time $\tau_\mu$ associated with the chemical potential difference is defined  via 
\begin{equation}
\frac{1}{\tau_\mu}\equiv \frac{\beta_0gn_{c0}}{\tau_{12}}\left(\frac{\frac{5}{2}\tilde P_0+2g\tilde n_0n_{c0}+\frac{2}{3}\tilde\gamma_0 g n_{c0}^2}{\frac{5}{2}\tilde\gamma_0\tilde P_0+\frac{2}{3} g n_{c0}^2}-1\right)
\equiv\frac{\beta_0gn_0}{\tau_{12}\sigma_H}.
\end{equation}

As discussed in Refs.~\cite{ T_Ara_PRA2009_F}, the liner response to the pulse perturbation can be described in terms of the density response function $\chi_{nn}({\bf q},\omega)$. We will thus calculate $\chi_{nn}({\bf q},\omega)$ by considering the external perturbation that excites plane wave $\delta U=\delta U_{{\bf q},\omega}e^{i({\bf q}\cdot{\bf r} -\omega t)}$.
Therefore we look for the plane-wave solutions
$\phi_{c,n}({\bf r},t) = \phi_{c,n,{\bf q},\omega} e^{i({\bf q}\cdot{\bf r} -\omega t)}$.
In this case, (\ref{greens_eq11}) reduces to
\begin{equation}
\delta \mu_{\rm diff} = {\tau_\mu \over 1-i\omega\tau_\mu}gn_{c0}\left ( \phi_c - {2\over 3}\phi_n \right ) q^2.
\label{greens_eq13}
\end{equation}
Substituting this result into (\ref{greens_eq9}) and (\ref{greens_eq10}), we are left with two coupled equations
for the superfluid and normal fluid velocity potentials: 
\begin{eqnarray}
M\omega^2 \phi_{c,{\bf q},\omega} &=& gn_{c0}
\Bigg ( 1 - {\sigma_H \over 1-i\omega\tau_\mu} \Bigg ) q^2 \phi_{c,{\bf q},\omega}\nonumber \\
&&+ 2g\tilde n_0 \left [ 1 + {\sigma_H \over 3
(1-i\omega\tau_\mu) } {n_{c0} \over \tilde n_0}
\right ] q^2 \phi_{n,{\bf q},\omega}-i\omega \delta U_{\bf{q},\omega},
\label{greens_eq14}
\end{eqnarray}
and
\begin{eqnarray}
M\omega^2 \phi_{n,{\bf q},\omega} = \Bigg \{ {5\tilde P_0 \over 3\tilde n_0} &+& 2g\tilde
n_0 \left [ 1- {2\sigma_H \over 9 (1-i\omega\tau_\mu) } {n_{c0}^2\over \tilde n_0^2} \right ]\Bigg \} q^2 
\phi_{n,{\bf q},\omega} \nonumber \\ 
&+& 2gn_{c0} \left [ 1 + {\sigma_H \over 3 (1-i\omega\tau_\mu) } {n_{c0} \over \tilde n_0}
\right ] q^2 \phi_{c,{\bf q},\omega} -i \omega \delta U_{\bf{q},\omega}.
\label{greens_eq15}
\end{eqnarray}

Taking the limit $\omega \tau_\mu \to 0$ of these coupled equations, we obtain
\begin{eqnarray}
M\omega^2 \phi_{c,{\bf q},\omega} &=& gn_{c0} ( 1 - \sigma_H ) q^2 \phi_{c,{\bf q},\omega}
+ 2g\tilde n_0 \left ( 1 + {\sigma_H n_{c0}
\over 3 \tilde n_0} \right ) q^2 \phi_{n,{\bf q},\omega}-i \omega \delta U_{\bf{q},\omega},
 \label{greens_eq16} \\
M\omega^2 \phi_{n,{\bf q},\omega} &=& 
\Bigg [ {5\tilde P_0 \over 3\tilde n_0} + 2g\tilde
n_0\left( 1 - {2\sigma_H n_{c0}^2\over 9\tilde n_0^2} \right) \Bigg ]
 q^2 \phi_{n,{\bf q},\omega}  \nonumber\\
&&{}{}{} + 2gn_{c0} \left ( 1 + {\sigma_H n_{c0} \over 3 \tilde n_0} \right ) q^2 \phi_{c,{\bf q},\omega}
-i \omega \delta U_{\bf{q},\omega}.
\label{greens_eq17}
\end{eqnarray}
It is useful to rewrite (\ref{greens_eq16}) and (\ref{greens_eq17}) in a
simple matrix form as
\begin{equation}
\left(
\begin{array}{cc} 
     \omega^2-v_2^2q^2 & -v_{21}^2q^2 \\
    -v_{12}^2q^2 & \omega^2-v_1^2q^2\\
      \end{array}\right)
\left(
   \begin{array}{c} 
     \phi_{c,{\bf q},\omega} \\
     \phi_{n,{\bf q},\omega} 
      \end{array}
  \right)
=-i \omega \delta U_{\bf{q},\omega}
\left(
   \begin{array}{c} 
    1  \\ 
1 
      \end{array}
  \right),
\label{greens_eq18}
\end{equation}
where we have introduced new velocities
\begin{eqnarray}
v_2^2&=&\frac{gn_{c0}}{M}(1-\sigma_H),~~
v_{21}^2=\frac{2g\tilde n_0}{M}
\left(1+\frac{\sigma_Hn_{c0}}{3\tilde n_0}\right), \nonumber \\
v_{12}^2&=&\frac{2g n_{c0}}{M}
\left(1+\frac{\sigma_Hn_{c0}}{3\tilde n_0}\right),~~
v_1^2=\frac{5\tilde P_0}{3M\tilde n_0}+\frac{2g\tilde n_0}{M}
\left( 1-\frac{2\sigma_Hn_{c0}^2}{9\tilde n_0^2} \right).
\label{greens_eq19}
\end{eqnarray}
We note that these new velocities are related to the first and second sound velocities $u_1$ and $u_2$
through
\begin{equation}
u_1^2+u_2^2=v_1^2+v_2^2,~~u_1^2u_2^2=v_1^2v_2^2-v_{12}^2v_{21}^2.
\label{greens_eq20}
\end{equation}
Solving (\ref{greens_eq18}), we obtain
\begin{equation}
\left(
   \begin{array}{c} 
     \phi_{c,{\bf q},\omega} \\
     \phi_{n,{\bf q},\omega} 
      \end{array}
  \right)
=-i \omega \delta U_{\bf{q},\omega}
\frac{1}{(\omega^2-u_1^2q^2)(\omega^2-u_2^2q^2)}
\left(
\begin{array}{c} 
     \omega^2-v_1^2q^2 + v_{21}^2q^2 \\
    v_{12}^2q^2 + \omega^2-v_2^2q^2\\
      \end{array}\right).
\label{greens_eq18b}
\end{equation}
Using (\ref{greens_eq13}) and Taking the limit $\omega \tau_\mu \to 0$,  (\ref{greens_eq5}) and (\ref{greens_eq1}) reduce to
\begin{eqnarray}
\left(
   \begin{array}{c} 
  \delta n_{c}\\
     \delta \tilde{n}
      \end{array}
  \right)=\frac{q^2}{-i\omega M} 
  \left(
   \begin{array}{cc} 
     n_{c0}(1+\sigma_H)&-\frac{2}{3}\sigma_Hn_{c0} \\
     -\sigma_H n_{c0}&\tilde{n}_0(1+\frac{2}{3}\frac{n_{c0}}{\tilde{n}_0})
      \end{array}
  \right)
  \left(
   \begin{array}{c} 
     \phi_{c,{\bf q},\omega} \\
     \phi_{n,{\bf q},\omega} 
      \end{array}
  \right).
  \label{eq_62}
\end{eqnarray}
Using the solution (\ref{greens_eq18b}) in the expression (\ref{eq_62}), we obtain 
\begin{eqnarray}
\left(
   \begin{array}{c} 
  \delta n_{c}\\
     \delta \tilde{n}
      \end{array}
  \right)&&=\frac{q^2 \delta U_{\bf{q},\omega}}{M(\omega^2-u_1^2q^2)(\omega^2-u_2^2q^2)}\nonumber \\
  && \times\left(
   \begin{array}{c} 
 n_{c0}(1+\frac{1}{3}\sigma_H)\omega^2+\{n_{c0}(1+\sigma_H)(-v_1^2+v_{21}^2)+\frac{2}{3}\sigma_Hn_{c0}(v_2^2-v_{12}^2)\}q^2 \\ \tilde{n}_0(1-\frac{1}{3}\sigma_H \frac{n_{c0}}{\tilde{n}_0})\omega^2+\{
 n_{c0}\sigma_H(v_1^2-v_{21}^2)+\tilde{n}_0(1+\frac{2}{3}\frac{n_{c0}}{\tilde{n}_0})(-v_2^2+v_{12}^2)\}q^2
          \end{array}  \right).
\end{eqnarray}
The solution can always be written in terms of the density response function, defined as   
\begin{eqnarray}
\left(
   \begin{array}{c} 
  \delta n_{c}\\
     \delta \tilde{n}
      \end{array}
       \right)
       =\delta U_{\bf{q},\omega}\left(
   \begin{array}{c} \chi_{n_c n } \\ \chi_{\tilde{n} n }
   \end{array}  \right),
\end{eqnarray}
with
\begin{eqnarray}
\chi_{n_c n }&=&\frac{q^2n_{c0}}{M}\frac{(1+\frac{1}{3}\sigma_H)\omega^2+\{(1+\sigma_H)(-v_1^2+v_{21}^2)+\frac{2}{3}\sigma_H(v_2^2-v_{12}^2)\}q^2}{(\omega^2-u_1^2q^2)(\omega^2-u_2^2q^2)}, \\
\chi_{\tilde{n} n }&=&\frac{q^2 \tilde{n}_0}{M}\frac{(1-\frac{1}{3}\sigma_H \frac{n_{c0}}{\tilde{n}_0})\omega^2+\{
 \frac{n_{c0}}{\tilde{n}_0}\sigma_H(v_1^2-v_{21}^2)+(1+\frac{2}{3}\frac{n_{c0}}{\tilde{n}_0})(-v_2^2+v_{12}^2)\}q^2}{(\omega^2-u_1^2q^2)(\omega^2-u_2^2q^2)}.
\end{eqnarray}

In the case of the sound propagation experiment,
a localized potential is applied at $t > 0$, while it is turned off at $t = 0$. This situation can be described as
$\delta U({\bf r},t)=\delta U(z)\theta(-t)$ \cite{T_Ara_PRA2009_F}.
In this case, the density fluctuations at $t>0$ is given by
\begin{eqnarray}
\delta n_c(z,t)=\frac{1}{2\pi^2}\int  dq\int d\omega \delta U(q)
\frac{\chi_{n_c n}^{\prime\prime}}{(w+i\eta)}e^{iqz-i\omega t} \ \ (t>0), \label{delta n_c} \\ 
\delta \tilde{n}(z,t)=\frac{1}{2\pi^2}\int  dq\int d\omega \delta U(q)
\frac{\chi_{\tilde{n} n}^{\prime\prime}}{(w+i\eta)}e^{iqz-i\omega t} \ \ (t>0), \label{delta n_ti} 
\end{eqnarray}
where $\chi_{n_c n}^{\prime\prime} (\mathbf{q},\omega)= {\rm{Im}}\chi_{n_c n}(\mathbf{q},\omega+i\eta)$ and $\chi_{\tilde{n} n}^{\prime\prime} (\mathbf{q},\omega)= {\rm{Im}}\chi_{\tilde{n} n}(\mathbf{q},\omega+i\eta)$. From (\ref{delta n_c}) and (\ref{delta n_ti}), we obtain 
\begin{eqnarray}
\delta n_c(z,t)&=&W_1^{n_c}\left[\delta U(z-u_1t)+\delta U(z+u_1t)\right]
+W_2^{n_c}\left[\delta U(z-u_2t)+\delta U(z+u_2t)\label{delta n_u}\right], \\ 
\delta \tilde{n}(z,t)&=&W_1^{\tilde{n}}\left[\delta U(z-u_1t)+\delta U(z+u_1t)\right]
+W_2^{\tilde{n}}\left[\delta U(z-u_2t)+\delta U(z+u_2t)\label{delta n_u}\right],
\label{den}
\end{eqnarray}
where the amplitudes of the  sound pulse are given by
\begin{eqnarray}
W_1^{n_c}&=&\frac{n_{c0}}{2M u_1^2}\frac{(1+\frac{1}{3}\sigma_H)u_1^2+(1+\sigma_H)(-v_1^2+v_{21}^2)+\frac{2}{3}\sigma_H(v_2^2-v_{12}^2)}{u_2^2-u_1^2}, \label{eq_71}\\
W_2^{n_c}&=&\frac{n_{c0}}{2M u_2^2}\frac{(1+\frac{1}{3}\sigma_H)u_2^2+(1+\sigma_H)(-v_1^2+v_{21}^2)+\frac{2}{3}\sigma_H(v_2^2-v_{12}^2)}{u_2^2-u_1^2}, \\ 
W_1^{\tilde{n}}&=&\frac{\tilde{n}_0}{2M u_1^2}\frac{(1-\frac{1}{3}\sigma_H \frac{n_{c0}}{\tilde{n}_0})u_1^2+
 \frac{n_{c0}}{\tilde{n}_0}\sigma_H(v_1^2-v_{21}^2)+(1+\frac{2}{3}\frac{n_{c0}}{\tilde{n}_0})(-v_2^2+v_{12}^2)}{u_2^2-u_1^2}, \\
 W_2^{\tilde{n}}&=&\frac{\tilde{n}_0}{2M u_2^2}\frac{(1-\frac{1}{3}\sigma_H \frac{n_{c0}}{\tilde{n}_0})u_1^2+
 \frac{n_{c0}}{\tilde{n}_0}\sigma_H(v_1^2-v_{21}^2)+(1+\frac{2}{3}\frac{n_{c0}}{\tilde{n}_0})(-v_2^2+v_{12}^2)}{u_2^2-u_1^2}.
 \label{eq_74}
\end{eqnarray}

\begin{figure}[htbp]
\includegraphics[height=2.in]{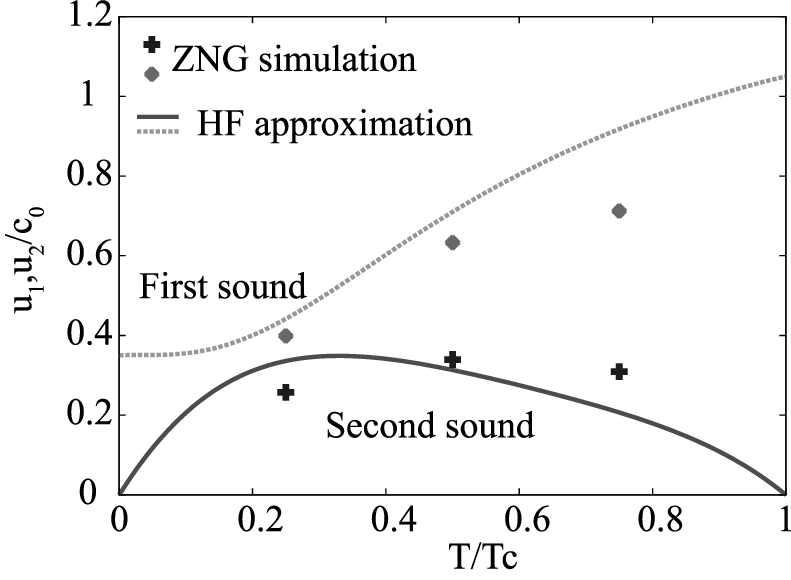}
 \caption{First and second sound velocities as a function of temperature.} 
  \label{Fig_vel}
\end{figure}

We estimate the interaction parameter for a uniform gas corresponding to the experiment of Ref.~\cite{E_second_Bose} from the average density of the trapped gas, and obtain $na^3\simeq 0.07$ and thus  
 $\frac{gn}{k_{\rm B}T_c^0}\simeq 0.5$,
were $T_c^0$ is the BEC transition temperature of an ideal Bose gas. 
In Fig.~\ref{Fig_vel}, we plot the first and second sound velocities as a function of temperature within the HF approximation, and compare with the sound velocities deduced from ZNG simulations discussed in the previous section.  
We emphasize that both first and second sound velocities obtained from ZNG simulation show good agreement with HF approximation. This confirms that our ZNG simulation well describe the two-fluid hydrodynamics.

\begin{figure}[htbp]
\includegraphics[height=2.in]{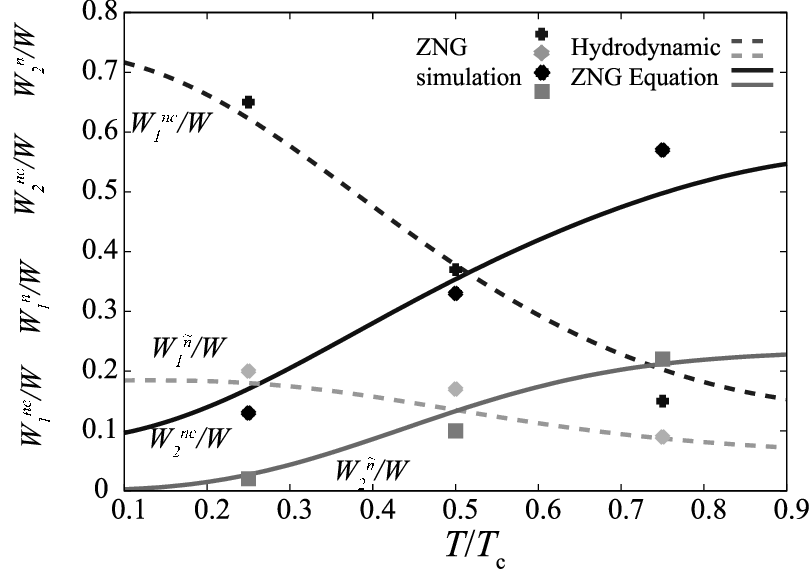}
 \caption{The first sound amplitude $W_1=W_1^{n_c}+W_1^{\tilde{n}}$ and second sound amplitude $W_2=W_2^{n_c}+W_2^{\tilde{n}}$ as a function of temperature, where $W=W_1+W_2=W_1^{n_c}+W_1^{\tilde{n}}+W_2^{n_c}+W_2^{\tilde{n}}$. Lines show the results from the self-consistent HF approximation. Data points are results from the simulation solving ZNG equations.} 
  \label{Fig_L_2}
\end{figure}

We now compare the pulse amplitude obtained from simulation results in the previous section with the ZNG hydrodynamic results for a uniform Bose gas.   
The amplitudes of first and second sound $W_i^{n_c}$ and $W_i^{\tilde n}$ are obtained by taking average of subtracting the unperturbed density profile from perturbed ones. 
In Fig.~\ref{Fig_L_2}, the first and second sound amplitudes for condensate and noncondensate components obtained by the simulation of ZNG equations (Eqs.~(\ref{GGP}) and (\ref{Bolt})) and the results (\ref{eq_71})-(\ref{eq_74}) which calculated by the linearized ZNG hydrodynamic equations. We see that the simulation results are consistent with the analytical results.  

\section{Conclusion}
In this paper, we have discussed sound propagation in Bose-condensed gases in a highly-elongated harmonic trap. In order to consider the situation of a highly-elongated harmonic trap, we derive quasi-1D ZNG equations. Using these equation, we show the several dynamical simulation with the same parameter as experiment on second sound. We showed that both first and second sound mode can be observed by a sudden modification of a trapping potential at intermediate temperatures.
We also found that the thermal density perturbations is so small that one cannot distinguish small density perturbations from signal-to-noise in the thermal cloud. 

We also derived expression for the pulse amplitude of condensate and noncondensate components in a uniform Bose gases using linearized ZNG hydrodynamic equations.
The first and second sound amplitude obtained by dynamical simulation are consistent with the results calculated by the linearized ZNG hydrodynamic equations. 
This also confirm that the system we considered in this paper is well described by the two-fluid hydrodynamics. The quasi-1D ZNG formalism developed in this paper is very useful in analyzing the finite temperature dynamics of highly-elongated Bose-condensed gases. In a separate paper, we will study the collective modes of a highly-elongated Bose gas at finite temperatures. We hope to stimulate further detailed  experimental examination on the dynamics of Bose condensed gases at finite temperatures. 
\section{Acknowledgmments}
We thank A. Griffin for valuable comments. Our program for dynamical simulation of sound propagation is based on the program developed by T. Inoue and S. Imai. 
E. A. is supported by a Grant-in-Aid from JSPS.
\appendix
\section{DERIVATION OF COLLISION INTEGRALS}
In this Appendix, we give a detailed derivation of the expressions for the collision integrals given by (\ref{C_12}) and (\ref{C_22}). We closely follow the approach of Refs.~\cite{T_ZNG_JLTP,B_GNZ}. 

According to the time dependent perturbation theory, the expectation value of an arbitrary operator $\hat O(t)$ made up of some
combination of non-condensate field operators can 
be expressed to first order in $\hat H'$ as
\begin{eqnarray}
\langle \hat O \rangle_t &=& {\rm Tr} \Bigg\{ \hat S_0(t,t_0)\hat O(t_0)\hat S_0(t,t_0)\nonumber \\
&&-\frac{i}{\hbar}\int_{t_0}^tdt^\prime \hat{S}^\dagger_0(t^\prime,t_0)[\hat{S}^\dagger_0(t,t^\prime)\hat O(t_0)\hat S_0(t,t^\prime),\hat H^\prime (t^\prime)]S_0(t^\prime,t_0)\Bigg\}.
\end{eqnarray}
The three-field correlation function is given by
\begin{eqnarray}
&&\langle \tilde\psi^\dagger_n(z,t) \tilde\psi_m(z,t) \tilde\psi_l(z,t)\rangle \nonumber \\
&&=-\frac{i}{\hbar}{\rm Tr} \hat \rho(t_0)\int_{t_0}^t dt^\prime \hat{S}^\dagger_0(t^\prime,t_0)[\hat{S}^\dagger_0(t,t^\prime)\tilde\psi^\dagger_n(z,t_0) \tilde\psi_m(z,t_0) \tilde\psi_l(z,t_0)\nonumber \\&&~~~~~~~~~~~~~~~~~~~~~~~~~~~~~~~~~~~~~~~~
\times \hat S_0(t,t^\prime),\hat H^\prime_1 (t^\prime)+\hat H^\prime_3 (t^\prime)]S_0(t^\prime,t_0)
\nonumber \\ &&
\equiv \langle \tilde\psi^\dagger_n(z,t) \tilde\psi_m(z,t) \tilde\psi_l(z,t) \rangle^{(1)}+ \langle \tilde\psi^\dagger_n(z,t) \tilde\psi_m(z,t) \tilde\psi_l(z,t)\rangle^{(3)},
\end{eqnarray}
where $\langle \cdots\rangle^{(i)}$ denote the contribution from $\hat H'_i$. 
In this case, contributions from the other terms in $\hat H'$ can be shown to vanish. 
The evaluation of correlation functions is facilitated by two key assumptions:
The effect of $H'_1(t')$ in the interval $t_0< t^\prime < t$ is essentially a collision process, which occurs on a time scale much shorter than all other time scales in the problem, and the hydrodynamic variables vary slowly in space and time.
It is sufficient to use
\begin{eqnarray}
&&n_c^{n}(z^\prime,t)\simeq n^{n}_c(z,t)\simeq  n_c(z,t),\ \tilde{n}^{n}(z^\prime,t)\simeq \tilde{n}^{n}(z,t), \nonumber \\
&& \theta(z^\prime,t)\simeq \theta (z,t)-\frac{\epsilon_c (z,t)}{\hbar}(t^\prime-t)+\frac{m{\bf v}_c(z,t)}{\hbar}(z^\prime-z),
\end{eqnarray}
and 
\begin{eqnarray}
\hat S_0(t,t^\prime)\simeq e^{-i\hat H_0(t)(t-t^\prime)}.
\end{eqnarray}
Introducing the Fourier transform of the non-condensate field
operators according to
\begin{eqnarray}
\tilde{\psi}_n(z,t_0)=\sum_{p_z}a_{p_z, n}e^{ip_zz}.
\end{eqnarray}
The contribution from $H'_1$ to the commutator is given by 
\begin{eqnarray}
&&[\hat{S}^\dagger_0(t,t^\prime)\tilde\psi^\dagger_n(z,t_0) \tilde\psi_m(z,t_0) \tilde\psi_l(z,t_0)\hat S_0(t,t^\prime),\hat H^\prime_1 (t^\prime)],\nonumber \\
&&=-2\sum_{n^\prime k^\prime l^\prime}g_{n^\prime n^\prime k^\prime l^\prime }\sum_{p_{z1}p_{z2}p_{z3}p_{z4}}e^{-i(p_{z1}-p_{z2}-p_{z3})z}e^{i(\tilde{\epsilon}_1^n-\tilde{\epsilon}_2^m-\tilde{\epsilon}_3^k)(t-t^\prime)}\int dz^\prime \tilde n^{n^\prime}(z^\prime,t )\Phi_{k^\prime}(z^\prime,t)\nonumber \\
&&[a^\dagger_{p_{z1}n}a_{p_{z2}m}a_{p_{z3}k},a^\dagger_{p_{z4}l^\prime}]
\nonumber \\
&&\simeq-2 \sum_{n^\prime l^\prime}g_{n^\prime n^\prime 0l^\prime }\tilde n^{n^\prime }(z,t )\sqrt{n_c}e^{i\theta(z,t)}
\sum_{p_{z1}p_{z2}p_{z3}p_{z4}}e^{-i(p_{zc}+p_{z1}-p_{z2}-p_{z3})z}e^{i(\epsilon_c+\tilde{\epsilon}_1^n-\tilde{\epsilon}_2^m-\tilde{\epsilon}_3^k)(t-t^\prime)}
\nonumber \\
&&\times \delta_{p_{z4}p_{zc}}[a^\dagger_{p_{z1}n}a_{p_{z2}m}\delta_{p_{z3}p_{z4}}\delta_{l^\prime,l}+a^\dagger_{p_{z1}n}a_{p_{z3}l}\delta_{p_{z2}p_{z4}}\delta_{l^\prime,m}],
\end{eqnarray}
where we have defined the condensate momentum $p_{zc}\equiv mv_c$.
We thus obtain
\begin{eqnarray}
\langle \tilde\psi^\dagger_n(z,t) \tilde\psi_m(z,t) \tilde\psi_l(z,t)\rangle_{(1)}
&=&-\frac{i}{\hbar}{\rm Tr} \hat \rho(t_0)\int_{t_0}^t dt^\prime \hat{S}^\dagger_0(t^\prime,t_0)[\hat{S}^\dagger_0(t,t^\prime)\tilde\psi^\dagger_n(z,t_0) \tilde\psi_m(z,t_0) \tilde\psi_l(z,t_0)\nonumber \\&&~~~~~~~~~~~~~~~~~~~~~~~~~~~~~~~~~~~~~~~~
\times \hat S_0(t,t^\prime),\hat H^\prime_1 (t^\prime)]S_0(t^\prime,t_0)
\nonumber \\
&=&2i \sum_{n^\prime l^\prime}g_{n^\prime n^\prime 0l^\prime }\tilde n^{n^\prime}(z,t )\sqrt{n_c}e^{i\theta(z,t)}\nonumber \\
&&
\sum_{p_{z1}p_{z2}p_{z3}p_{z4}}e^{-i(p_{zc}+p_{z1}-p_{z2}-p_{z3})z/\hbar }\delta_{p_{z4}p_{zc}}
\nonumber \\
&&\times \int_{t_0}^{t}dt^\prime e^{i(\epsilon_c+\tilde{\epsilon}_1^n-\tilde{\epsilon}_2^m-\tilde{\epsilon}_3^k)(t-t^\prime)/\hbar }[\langle a^\dagger_{p_{z1}n}a_{p_{z2}m}\rangle\delta_{p_{z3}p_{z4}}\delta_{l^\prime,l}\nonumber \\
&&
+\langle a^\dagger_{p_{z1}n}a_{p_{z3}l}\rangle \delta_{p_{z2}p_{z4}}\delta_{l^\prime,m}],
\end{eqnarray}
where
\begin{eqnarray}
\langle a^\dagger_{p_{z1}n}a_{p_{z2}m} \rangle_{t^\prime}={\rm Tr}\hat \rho(t_0)\hat S_0^\dagger (t^\prime,t_0)a^\dagger_{p_{z1}n}a_{p_{z2}m}S_0(t^\prime,t_0)\simeq e^{i(\tilde\epsilon_1-\tilde\epsilon_2)(t^\prime-t_0)}\langle a^\dagger_{p_{z1}n}a_{p_{z1}m}\rangle_{t_0}.
\end{eqnarray}
We now assume that the initial statistical density matrix $\hat \rho(t_0)$ has the form appropriate for the HF Hamiltonian
\begin{eqnarray}
\langle a^\dagger_{p_{z1}n}a_{p_{z1}m}\rangle_{t_0} \simeq \delta_{p_{z1},p_{z2}}\delta_{nm}f^n(p_{z1},z,t),
\end{eqnarray}
where we have assume that off-diagonal contribution $n\neq m$ is neglecting small.
In fact, in equilibrium, $f_{nm}(n\neq m)<10^{-5}$ and $\sum_{n\neq m} f_{nm}<10^{-3}$.     
Similarly, we can obtain
\begin{eqnarray}
&&\langle\tilde\psi^\dagger_n(z,t)\tilde\psi_m(z,t)\tilde\psi_l(z,t) \rangle_{(3)}
\nonumber \\ 
&=&-\frac{i}{\hbar}{\rm Tr} \hat \rho(t_0)\int_{t_0}^t dt^\prime \hat{S}^\dagger_0(t^\prime,t_0)[\hat{S}^\dagger_0(t,t^\prime)\tilde\psi^\dagger_n(z,t_0) \tilde\psi_m(z,t_0) \tilde\psi_l(z,t_0)
\times \hat S_0(t,t^\prime),\hat H^\prime_3 (t^\prime)]S_0(t^\prime,t_0)
\nonumber \\
&=&-2\sum_{n^\prime m^\prime k^\prime} g_{n^\prime m^\prime k^\prime 0}n_c^{1/2}e^{i\theta}\sum_{ p_{z1},p_{z2}p_{z3}}e^{i(p_{z2}+p_{z3}-p_{z1}-p_{zc})z}\nonumber \\ &&
\int_{t_0}^t dt^\prime e^{i(\epsilon_c+\tilde{\epsilon}_1^n-\tilde{\epsilon}_2^m-\tilde{\epsilon}_3^k)(t-t^\prime)} \sum_{ p_{z1}^\prime,p_{z2}^\prime p_{z3}^\prime}
\delta_{p_{cz}+p_{z1}^\prime,p_{z2}^\prime+p_{z3}^\prime}
\nonumber \\
&&\times [\delta_{p_2,p_2^\prime}\delta_{p_3,p_3^\prime}\delta_{mm^\prime}\delta_{kk^\prime}\langle a^\dagger_{p_{z1}n}a_{p_{z1}^\prime n^\prime}\rangle_{t^\prime}+\delta_{p_2,p_2^\prime}\delta_{mm^\prime}\langle a^\dagger_{p_{z1}n}a_{p_{z1}^\prime n^\prime}\rangle_{t^\prime}\langle a^\dagger_{p_{z3}k}a_{p_{z3}^\prime k^\prime}\rangle_{t^\prime}
\nonumber \\
&&
+\delta_{p_3,p_3^\prime}\delta_{ll^\prime}\langle a^\dagger_{p_{z1}n}a_{p_{z1}^\prime n^\prime}\rangle_{t^\prime}\langle a^\dagger_{p_{z2}m}a_{p_{z2}^\prime m^\prime}\rangle_{t^\prime}-\delta_{p_1,p_1^\prime}\delta_{nn^\prime}\langle a^\dagger_{p_{z2}m}a_{p_{z2}^\prime m^\prime}\rangle_{t^\prime}\langle a^\dagger_{p_{z3}k}a_{p_{z3}^\prime k^\prime}\rangle_{t^\prime}\nonumber \\
&&
+\delta_{p_2,p_2^\prime}\delta_{mm^\prime}\langle a^\dagger_{p_{z1}n}a_{p_{z3} k}\rangle_{t^\prime}\langle a^\dagger_{p_{z3}^\prime k^\prime} a_{p_{z1}^\prime n^\prime}\rangle_{t^\prime}+\delta_{p_1,p_1^\prime}\delta_{n,n^\prime}\langle a^\dagger_{p_{z1}n}a_{p_{z2} m}\rangle_{t^\prime}\langle a^\dagger_{p_{z2}^\prime m^\prime} a_{p_{z1}^\prime n^\prime}\rangle_{t^\prime}.
\end{eqnarray}
The last two terms reduce to 
\begin{eqnarray}
&& \sum_{ p_{z1}^\prime,p_{z2}^\prime p_{z3}^\prime}\delta_{p_2,p_2^\prime}\delta_{p_{cz}+p_{z1}^\prime,p_{z2}^\prime+p_{z3}^\prime}(\delta_{mm^\prime}\langle a^\dagger_{p_{z1}n}a_{p_{z3} k}\rangle_{t^\prime}\langle a^\dagger_{p_{z3}^\prime k^\prime} a_{p_{z1}^\prime n^\prime}\rangle_{t^\prime}\nonumber \\&&~~~~~~~~~~~~~
+\delta_{p_1,p_1^\prime}\delta_{n,n^\prime}\langle a^\dagger_{p_{z1}n}a_{p_{z2} m}\rangle_{t^\prime}\langle a^\dagger_{p_{z2}^\prime m^\prime} a_{p_{z1}^\prime n^\prime}\rangle_{t^\prime})\nonumber \\
~~~~&&=\tilde{n}^{n^\prime}  \sum_{p_{z2^\prime}}\delta_{p_{zc},p_{z2}^\prime}\left(\delta_{p_{z2},p_{z2}^\prime}\delta_{m,m^\prime}\langle a^\dagger_{p_{z1}n}a_{p_{z3} k}\rangle_t+\delta_{p_{z3},p_{z2}^\prime}\delta_{k,m^\prime}\langle a^\dagger_{p_{z1}n}a_{p_{z2} m}\rangle_t\right).
\end{eqnarray}
We now see that the last two terms in this equation exactly cancel the contribution from $\hat H_1^\prime$. We thus obtain
\begin{eqnarray}
\langle\tilde\psi^\dagger_n(z,t)\tilde\psi_m(z,t)\tilde\psi_k(z,t) \rangle_t
&=&-i 2 \pi g_{nmk0}n_c^{1/2}e^{i\theta}
\nonumber \\&&\sum_{p_{z1},p_{z2},p_{z3}}\Bigg[\delta(\epsilon_c+\tilde{\epsilon}_1^n-\tilde{\epsilon}_2^m-\tilde{\epsilon}_3^k)+\frac{i}{\pi}P\frac{1}{(\epsilon_c+\tilde{\epsilon}_1^n-\tilde{\epsilon}_2^m-\tilde{\epsilon}_3^k)}\Bigg]\nonumber \\&&
\delta_{p_{zc}+p_{z1},p_{z2}+p_{z3}}[f_1^n(1+f_2^m)(1+f_3^k)-(1+f_1^n)f_2^mf_3^k],
\label{A_12}
\end{eqnarray}
with $f_i^l=f^l(p_{zi},z_i,t)$. In
addition, we have treated the system as locally homogeneous, with the
consequence that the local HF single-particle energies,
$\tilde{\epsilon}^l=\frac{p_{z}^2}{2M}+E_l+U^{l}(z,t)$.
Using (\ref{A_12}), we obtain the $C_{12}$ collision integral 
\begin{eqnarray}
C_{12}[f_{n}]&\equiv&-i{\rm Tr} \rho(t,t_0)\left[\hat{f}_{n}(p_z,z,t_0),\hat H_{3}^\prime(t)\right]\nonumber\\ & 
\simeq&-i\sum_{n^\prime m^\prime k^\prime}g_{n^\prime m^\prime k^\prime 0}n_c^{1/2}e^{-i\theta}\sum_{q_z}\sum_{p_{z1},p_{z2},p_{z3}}e^{iq_zz}
\delta(p_{cz}+p_{z1},p_{z2}+p_{z3})e^{ip_{zc}z}\nonumber\\ &&[
\delta_{p_{z1},p_{z}+q_{z}/2}\delta_{mn^\prime}\langle a^\dagger_{p_{z}-q_{z}/2n} a_{p_{z2}m^\prime}a_{p_{z3} k^\prime}\rangle_t
-\delta_{p_{z2},p_{z}-q_{z}/2}\delta_{nm^\prime}\langle a^\dagger_{p_{z1}n} a_{p_{z}-q_{z}/2 m}a_{p_{z3} k^\prime}\rangle_t\nonumber\\ &&
-\delta_{p_{z3},p_{z}-q_{z}/2}\delta_{nk^\prime}\langle a^\dagger_{p_{z1}n} a_{p_{z2} m^\prime}a_{p_{z}-q_{z}/2 m}\rangle_t -h.c.]\nonumber \\
&=&4\pi \sum_{n^\prime m^\prime k^\prime}g_{n^\prime m^\prime k^\prime 0}^2n_c 
\sum_{p_{z1},p_{z2},p_{z3}}\left[\delta(\epsilon_c+\tilde{\epsilon}_1^{n'}-\tilde{\epsilon}_2^{m'}-\tilde{\epsilon}_3^{k'})\right]
\nonumber \\&&
\times\delta_{p_{zc}+p_{z1},p_{z2}+p_{z3}}
(\delta_{p_z,p_{z1}}\delta_{nn'}+\delta_{p_{z},p_{z2}}\delta_{nm'}-\delta_{p_{z},p_{z3}}\delta_{nl'})
\nonumber \\ && \times\left[(1+f_1^{n'})f_2^{m'}f_3^{k'}-f_1^{n'}(1+f_2^{m'})(1+f_3^{k'})\right].
\end{eqnarray}
Similarly, we can obtain the expression for $C_{22}$ collision term, which is the contribution from the $\hat H_4^\prime$ perturbation.
\begin{eqnarray}
C_{22}[f_{n}]&\equiv&-i{\rm Tr}\tilde{\rho}(t,t_0)[\hat f(p_z,z,t_0),\hat H_4^\prime(t)]\nonumber \\&
\simeq& \frac{-i}{2}\sum_{n^\prime m^\prime k^\prime l^\prime}\sum_{q_z}\sum_{p_{z1},p_{z2},p_{z3}}e^{iq_zz}\delta(p_{z1}+p_{z2},p_{z3}+p_{z4})
\nonumber \\&&
\times[\delta_{p_{z1},p_{z}+q_{z}/2}\delta_{mn^\prime}\langle a^\dagger_{p_{z}-q_{z}/2n} a^\dagger_{p_{z2} m^\prime}a_{p_{z3} k^\prime}a_{p_{z4} l^\prime}\rangle_t \nonumber \\&&
+\delta_{p_{z2},p_{z}+q_{z}/2}\delta_{mm^\prime}\langle a^\dagger_{p_{z}-q_{z}/2n} a^\dagger_{p_{z1} n^\prime}a_{p_{z3} k^\prime}a_{p_{z4} l^\prime}\rangle_t\nonumber \\&&
-\delta_{p_{z3},p_{z}-q_{z}/2}\delta_{nk^\prime}\langle a^\dagger_{p_{z1}n^\prime} a^\dagger_{p_{z2} m^\prime}a_{p_{z4} l^\prime}a_{p_{z}+q_{z}/2 m}\rangle_t\nonumber \\&&
-\delta_{p_{z4},p_{z}-q_{z}/2}\delta_{nl^\prime}\langle a^\dagger_{p_{z1}n^\prime} a^\dagger_{p_{z2} m^\prime}a_{p_{z3} k^\prime}a_{p_{z}+q_{z}/2 m}\rangle_t
]
\label{A24}.
\end{eqnarray}
Using Wick's theorem, we
obtain the relevant contribution
\begin{eqnarray}
\langle a^\dagger_{p_{z1}n} a^\dagger_{p_{z2}m}a_{p_{z3} k}a_{p_{z4}l}\rangle_t&=&-2\pi i g_{nmkl}\delta(\tilde \epsilon_1^n+\tilde \epsilon_2^m-\tilde \epsilon_3^k-\tilde \epsilon_4^l)\delta_{p_{z1}+p_{z2},p_{z3}+p_{z4}}\nonumber \\&&
\times
[f_1^nf_2^m(1+f_3^k)(1+f_4^l)-(1+f_1^n)(1+f_2^m)f_3^kf_4^l]\label{A26}.
\end{eqnarray}
Inserting (\ref{A26}) into (\ref{A24}), we obtain  
\begin{eqnarray}
C_{22}[f_{n}]&=&\pi \sum_{n'm'k'l'}g_{n'm'k'l'}^2
\sum_{p_{z1},p_{z2},p_{z3},p_{z4}}\left[\delta(\tilde{\epsilon}_1^{n'}-\tilde{\epsilon}_2^{k'}-\tilde{\epsilon}_3^{m'}-\tilde{\epsilon}_4^{l'})\right]
\nonumber \\&&
\times \delta_{p_{z1}+p_{z2},p_{z3}+p_{z4}}
(\delta_{p_z,p_{z1}}\delta_{nn'}+\delta_{p_{z},p_{z2}}\delta_{nk'}-\delta_{p_{z},p_{z3}}\delta_{nm'}-\delta_{p_{z},p_{z4}}\delta_{nl'})\nonumber \\&&
\times \left[f_1^{n'}f_2^{k'}(1+f_3^{m'})(1+f_4^{l'})-(1+f_1^{n'})(1+f_2^{k'})f_3^{m'}f_4^{l'}\right].\nonumber \\
\end{eqnarray}
\section{Numerical methods}
In this section we discuss solution of the collisionless
Boltzmann equation using N-body simulations.
The effect of collisions is dealt with later. It is generally very difficult to solve using standard methods for treating partial differential equations. An alternative approach used extensively in the literature is to represent the phase-space density $f_{n}(p_z, z,t)$
by a cloud of discrete test particles. The momentum and position of each particle in an external potential $U^{n}(z,t)$ is then evolved according to Newton's equations. 
The $i$th test particle has variables $\{z_i(t),p_i(t),n_i(t)\}$. Test particles keep motion in one dimension along $z$-axis. 

The phase-space distribution for
this situation is given by
\begin{eqnarray}
f_{n}(p_z,z,t)=\frac{\tilde N}{\tilde N_{tp}}2\pi\hbar \sum_{i=0}^{\tilde N_{tp}}\delta(z-z_i)\delta(p_z-{p_z}_i)\delta_{n,{n}_i}.
\end{eqnarray}
where the weighting factor is fixed by the requirement that
the phase-space distribution is normalized to the number of
physical atoms, $\tilde N$, with $\tilde N \hbar =\sum_{n}\int dz d p_zf_{n}$.
By using a sufficiently large number of test particles, $\tilde{N}_T$, a reasonable approximation to the continuous phase-space distribution is obtained.
Note that the number of test and physical particles is
not necessarily equal. In fact, for a relatively small number
of physical atoms it is essential to simulate more test particles in order to minimize the effects of a discrete particle description. 

The time evolution of $f_{n}(p_z,z,t)$ is determined by the time-dependent potential and momentum variables of each test particle, given by the equations 

\begin{eqnarray}
\frac{dz_i}{dt}&=&\frac{p_{z_i}(t)}{M},\nonumber \\
\frac{p_{z_i}(t)}{dt}&=&-\frac{d}{dz}U^{n_i}(z,t)|_{z=z_i}.
\end{eqnarray}
The $n_i(t)$ should change by collision process.
The collisions are treated in a similar manner to Ref.~\cite{T_Jacson_Zaremba,B_GNZ} except calculation of angle.
The $i$th test particle has the index of radial direction $n_i$ instead of angle. 

The phase-space variables are updated by advancing the
position and momentum of each particle at discrete time
steps $\Delta t$. 
Symplectic integrators are used extensively in molecular
dynamics (MD) simulations since they possess several desirable
properties, such as conservation of phase-space volume
and of energy over a long period. We use a second-order symplectic
integrator in our calculations, which is the classical
analog of the split-operator method discussed earlier. To
show this, it is convenient to work within the Lie formalism. Consider the classical Hamiltonian for a single particle, $H_i=\frac{p_{zi}^2}{2M}+V_i(z_i)$. The evolution of its phase-space coordinates
$Z_i=(p_{zi},z_i)$ is then determined by the equation
\begin{eqnarray}
\frac{dZ_i}{dt}=\{Z_i,H_i\}\equiv -i{\cal L}Z_i,
\end{eqnarray}
where $\{F,G\}=\sum_j \left(\partial_{z_j}F \partial_{p_{zi}}G- \partial_{p_{zi}}F\partial_{z_j}G\right)$ is the Poisson
bracket and ${\cal L}$ is the Liouville operator.
One can then write the time evolution $Z$ as 
\begin{eqnarray}
Z(t+\Delta t)=e^{-i{\cal L}\Delta t}Z(t).
\end{eqnarray}
Splitting the Hamiltonian into potential and kinetic terms, the effect of the classical operator in the simulations is to update the
particle positions and velocities in three steps
\begin{eqnarray}
&&\tilde z_{i}=z_{i}+\frac{1}{2}\Delta t v_i(t),\nonumber \\ &&
v_{i}(t+\Delta t)=v_{i}-\frac{\Delta t}{M}\frac{\partial}{\partial z_i} U^{n_i}(\tilde z_{i}), \nonumber \\ &&
z_{i}(t+\Delta t)=\tilde z_{i}+\frac{1}{2}\Delta t v_i (t+\Delta t).
\end{eqnarray}

The effective potential $U$ is determined self-consistently
as the system evolves in time, and includes the condensate
mean-field and the mean-field generated by the
thermal cloud. 
The latter is in general much weaker than the condensate mean-field due to the larger spatial extent of the thermal cloud.
Nevertheless, it is important to include this term in order to
ensure the conservation of the total energy of the system. 
Although the calculation of the condensate mean-field is
straightforward, the use of discrete particles with a contact
interatomic potential creates a problem in determining the
noncondensate mean-field. 
Taken literally, the mean-field consists of a series of delta peaks 
\begin{eqnarray}
\bar U^{n}_T(z,t)=\frac{\tilde N}{\tilde N_{tp}}2 \sum_{k,l}g_{nnkl}\sum_{i=0}^{\tilde N_{tp}}\delta(z-z_i)\delta_{n,{n}_i}
\equiv 2g^{n}\tilde{n}^{n}_T(z,t).
\end{eqnarray}
This expression clearly cannot be used as it is to generate the
forces acting on the test particles that are required in the MD
simulation.  We generate a smooth thermal cloud density by performing a convolution
with a sampling (or smoothening) function $S(z)$ which is
normalized to unity. In particular, we define
\begin{eqnarray}
\tilde{U}^{n}_{S}(z,t)\equiv \int dz' S(z-z^\prime) \tilde{U}^{n}_{T}(z',t)
=\frac{\tilde N}{\tilde N_{tp}}2 \sum_{k,l}g_{nnkl}\sum_{i=0}^{\tilde N_{tp}}S(z-z_i)\delta_{n,{n}_i},
\end{eqnarray}
where we choose $S(z)\sim e^{-z^2/\eta^2}$, i.e., an isotropic Gaussian sampling function of width $\eta$.

We proceed by making use of a FFT. First, each particle in the
ensemble is assigned to points on the 1D Cartesian grid using
a cloud-in-cell method. We now consider  a particle at position $z$, between two grid points at $z_k$ and $z_{k+1}$. The particle is assigned
to both points with weightings $(1-\alpha)$ and $\alpha$, respectively,
where $\alpha=({z-z_{k}})/({z_{k+1}-z_{k}})$. This can be viewed as
a more sophisticated binning procedure in that it takes into
account the actual positions of particles within the cells. We then convolve the cloud-in-cell density with the sampling function by Fourier transforming it and then multiplying it by the analytic FT
of the sampling function. An inverse FFT then generates the
sampled potential.  This potential is used directly
in the GP evolution, while the forces on the test particles
are obtained by taking a numerical derivative and interpolating
to the positions of the particles. We have also checked
that small variations of $\eta$ about the value chosen to do the
simulations have little effect on our final results.

Probabilities for either $C_{22}$ or $C_{12}$ collisions are calculated
in a way which is consistent with a Monte Carlo sampling
of the collision integrals, as discussed below.
We first give details for the $C_{22}$ integral, which physically
corresponds to scattering of two thermal particles into
two final thermal states. Hence the process conserves the
number of thermal atoms $\int dp_z/(2\pi \hbar)C_{22}=0$.  
We are interested
in the mean collision rate at a point $z$, which is given by
\begin{eqnarray}
\Gamma_{22}^{\rm out}[f_{n}]&=&\int \frac{dp_z}{2\pi \hbar}C_{22}^{\rm out}[f_{n}],
\end{eqnarray}
where 
\begin{eqnarray}
C_{22}^{\rm out}[f_{n}]&\equiv& \sum_{n'm'k'l'}\frac{g_{n'm'k'l'}^2}{4\pi \hbar ^2}
\int dp_{z2}dp_{z3}dp_{z4}\left[\delta(\tilde{\epsilon}_1^{n'}-\tilde{\epsilon}_2^{k'}-\tilde{\epsilon}_3^{m'}-\tilde{\epsilon}_4^{l'})\right]
\nonumber \\&&
\times \delta({p_{z}+p_{z2}-p_{z3}-p_{z4}})
(\delta_{nn'}+\delta_{nk'}-\delta_{nm'}-\delta_{nl'})\nonumber \\&&
\times f_1^{n'}f_2^{k'}(1+f_3^{m'})(1+f_4^{l'}).
\label{C_22_1}
\end{eqnarray}
We now write the required local collision rate as 
\begin{eqnarray}
&&\Gamma_{22}^{\rm out}[f_{n}]= \sum_{n'm'k'l'}\int  \frac{dp_{z1}}{2\pi \hbar}\int  \frac{dp_{z2}}{2\pi \hbar}f^{n'}(p_{z1})f^{k'}(p_{z2})g^{m'l'}_{n'k'}(p_{z1},p_{z2})
\nonumber \\ &&
=\sum_{kl} \int dp_z w^{n'k'}(p_z)g^{m'l'}_{n'k'}(p_z),
\end{eqnarray}
where $p_z$ is a point in two-dimensional momentum space and
the factor 
\begin{eqnarray}
w^{n'k'}(p_z)=f^{n'}(p_{z1})f^{k'}(p_{z2})/(2\pi \hbar)^2,
\end{eqnarray}
 is considered as a weight function. We denote the maximum value
of $w^{n'k'}(p_z)$ by $w^{n'k'}_{\rm max}$ and define the domain on which the integrand is nonzero by [$-p_{z{\rm max}}/2,p_{z{\rm max}}/2$] for each momentum component. Choosing a point $p_{zi}$
at random in the hypervolume $(p_{z{\rm max}})^2$, and a
random number $R_i$ uniformly distributed on [$0, w^{n'k'}_{\rm max}$] the point $p_{zi}$ is is accepted if $R_i<w^{n'k'}_{\rm max}$ and the quantity $g^{m'l'}_{n'k'}(p_{zi})$ is accumulated. The value of the integral is then given approximately as
\begin{eqnarray}
\Gamma_{22}^{\rm out}[f_{n}]\simeq  \sum_{n'm'k'l'}(p_{z {\rm max}})^2w^{n'k'}_{\rm max}
\frac{1}{N}{\sum_i}'g^{m'l'}_{n'k'}(p_{zi}),
\end{eqnarray}
where $N$ is the number of random $p_{zi}$ points chosen and the
prime on the summation includes only those points for which $R_i<w^{nk}(p_{iz})$. For $g^{ml}_{nk}$, the integral is simply
\begin{eqnarray}
{\tilde{n}^{n}(z)}{\tilde{n}^{k}(z)}=(p_{z {\rm max}})^2w^{nk}_{\rm max}
\frac{N_s^{nk}}{N},
\end{eqnarray}
where $N_s^{nk}$ is the total number of points accepted, and 
\begin{eqnarray}
\Gamma_{22}^{\rm out}[f_{n}]=  \sum_{n'm'k'l'}{\tilde{n}^{n'}(z)}{\tilde{n}^{k'}(z)} \frac{1}{N_s^{n'k'}}{\sum_i}'g^{m'l'}_{n'k'}(p_{zi}).
\end{eqnarray}
The sample of $N_s^{nk}$ points accepted consists of $N_s^{nk}p_{z1}$ values and $N_s^{nk}p_{z2}$ values, each of which is distributed according to $f^n(p_{z1})$ and $f^k(p_{z2})$. This set of $2Ns^{nk} p$ values can be identified with $N_{\rm cell}$
test particles in a cell of volume $\Delta z$. If this set is to be
representative of the local density, we must have 
\begin{eqnarray}
\tilde n^{k}(z)=\frac{N_{\rm cell}}{\Delta z}=\frac{2N^{nk}_s}{\Delta z}.
\end{eqnarray}
With this identification,
\begin{eqnarray}
\Delta z\Gamma_{22}^{\rm out}[f_{n}]= 2 \sum_{n'm'k'l'}{\tilde{n}^{n'}(z)}{\sum_i}'g^{m'l'}_{n'k'}(p_{z1}^i,p_{z2}^i).
\end{eqnarray}
In other words, the collision rate can be estimated by sampling
the test particles in the cell $\Delta z$ in paris. 

For our purposes it is convenient to express the integral in
terms of new momentum variables $(p_{z0},p'_{z0})$ and $(p''_z,p'''_z)$: $p_{z1,z2}=(p_{z0}\pm p''_z)/\sqrt{2}$ and $p_{z3,z4}=(p'_{z0}\pm p'''_z)/\sqrt{2}$. $p_{z0}$ and $p''$ are proportional to the center-of-mass and relative momenta,
respectively, of the incoming 1 and 2 particles.
The momentum and energy delta functions reduce to 
 \begin{eqnarray}
 \delta(\tilde{\epsilon}_1^n+\tilde{\epsilon}_2^k-\tilde{\epsilon}_3^m-\tilde{\epsilon}_4^l)
\delta({p_{z1}+p_{z2}-p_{z3}-p_{z4}})
=\frac{M}{\sqrt{2}}\delta(p_{z0}-p'_{z0})\delta(p''^2_z-p'''^2_z+\tilde E_{nk}^{ml}),
\end{eqnarray}
with $\tilde E_{nk}^{ml}=2M(E_n+E_k-E_l-E_m)$. Integrating over $p_{z0}^\prime$ and $p'''_z$, we obtain
 \begin{eqnarray}
 \Gamma^{\rm out}_{22}[f_{n}]= \sum_{n'm'k'l'}\frac{g_{n'm'k'l'}^2M}{4\pi\hbar^2}
 (\delta_{nn'}+\delta_{nk'}-\delta_{nm'}-\delta_{nl'})
 \nonumber \\ 
\int dp_{z1}f_1^{n'}\int dp_{z2}f_2^{k'}
|{p_{z1}-p_{z2}}|(1+f_3^{m'})(1+f_4^{l'}).
\end{eqnarray}
where $p_{z3,z4}=p_{z0}\pm\sqrt{p''^2_z+\tilde E_{nk}^{ml}}$
Calculation of the rate therefore involves integrals over all possible initial states and all radial states.
Inserting the
explicit form of $g$ for $\Gamma_{22}$ collision rate, we have
\begin{eqnarray}
\Delta z \Gamma^{\rm out}_{22}[f_{n}]=\sum_{kl}2\pi Mg_{nmkl}^2\sum_{(ij)}\tilde n^{k}(z)|p_{zi}-p_{zj}|(1+f_3^m)(1+f_4^l),
\label{gamma_22^out}
\end{eqnarray}
where the sum is now taken over pairs of test particles. This
expression allows us to define the probability $P_{i j}^{22}$ that a pair of atoms $(i j)$ in the cell suffers a collision in a time interval $\Delta t$,
\begin{eqnarray}
P^{22}_{ij}[f_{n}]=\sum_{n'm'k'l'}\pi Mg_{n'm'k'l'}^2
(\delta_{nn'}+\delta_{nk'}-\delta_{nm'}-\delta_{nl'})
\nonumber \\ 
\tilde n^{k'}(z)|p_{zi}-p_{zj}|(1+f_3^{m'})(1+f_4^{l'}).
\end{eqnarray}
Selecting atoms in pairs from each cell and assigning them a collision probability $P_{i j}^{22}$ allows us to simulate the effect of
collisions in a way which is consistent with the Boltzmann
collision integral.

We treat $C_{12}$ collisions somewhat differently. First, we
note that the total rate of change of the number of thermal
atoms per unit volume due to these collisions is

\begin{eqnarray}
\int \frac{dp_z}{2\pi\hbar}C_{12}[f_n]&=&\sum_{n'm'k'}\frac{g_{n'm'k'0}^2n_c}{\pi\hbar^3}
 \int dp_{z2} \int dp_{z3}\int dp_{z4} \delta(\epsilon_c+\tilde{\epsilon}_2^{n'}-\tilde{\epsilon}_3^{m'}-\tilde{\epsilon}_4^{k'})\delta({p_{cz}+p_{z2}-p_{z3}-p_{z4}})
\nonumber \\ && \times(\delta_{nn'}-\delta_{nm'}-\delta_{nk'})\left[f_2^{n'}(1+f_3^{m'})(1+f_4^{k'})-(1+f_2^{n'})f_3^{m'}f_4^{k'}\right]
\nonumber \\ &&
\equiv \Gamma^{\rm out}_{12}[f_{n}]-\Gamma^{\rm in}_{12}[f_{n}].
\end{eqnarray}
According to this definition, using same transformation as in (\ref{gamma_22^out})
\begin{eqnarray}
\Gamma^{\rm out}_{12}[f_{n}]&=&\sum_{n'm'k'}\frac{g_{n'm'k'0}^2n_c}{\pi\hbar^3}
 \int dp_{z2} \int dp_{z3}\int dp_{z4} \delta(\epsilon_c+\tilde{\epsilon}_2^n-\tilde{\epsilon}_3^m-\tilde{\epsilon}_4^k)
\nonumber \\ && \times (\delta_{nn'}-\delta_{nm'}-\delta_{nk'})\delta({p_{cz}+p_{z2}-p_{z3}-p_{z4}}) f_2^{n'}(1+f_3^{m'})(1+f_4^{k'})
\nonumber \\ 
&=&\sum_{n'm'k'}\int dp_{z2} \frac{g_{n'm'k'0}^2Mn_c}{\pi\hbar^3} f_2^{n'}p_z^{\rm out} (1+f_3^{m'})(1+f_4^{k'})(\delta_{nn'}-\delta_{nm'}-\delta_{nk'}),
\end{eqnarray}
where $p_z^{\rm out}=\sqrt{|p_{cz}-p_{z2}|^2-4M(U^{n}+E_n-\mu_c)}$. 
This rate can be estimated by writing
\begin{eqnarray}
\Gamma_{12}^{\rm out}[f_{n}]=\sum_{n'm'k'} \int dp_{z2} w^{n'}(p_{z2})g^{m'}_{n'k'}(p_{z2})(\delta_{nn'}-\delta_{nm'}-\delta_{nk'}),
\end{eqnarray}
where $w^{n}(p_{z2})=f^n(p_{z2})/(2\pi \hbar)$ and $g^{m}_{nk}(p_{z2})$ is the remaining part of the
integrand.
A Monte Carlo sampling of the integral leads to the estimate 
\begin{eqnarray}
\Delta z \Gamma_{12}^{\rm out}[f_{n}] \simeq \sum_{n'm'k'}\sum_{i=1}^{N_s}g^{m}_{nk}(p_{zi}) 
(\delta_{nn'}-\delta_{nm'}-\delta_{nk'}),
\end{eqnarray}
where $N_s$ represents the number of atoms in the cell of volume
$\Delta z$. The probability of an atom in the cell suffering
this kind of collision in the time interval $\Delta t$ is therefore
\begin{eqnarray}
&&P_{i_{nmk}}^{\rm out}[f_{n}]=g^{m}_{nk}(p_{zi}) \Delta t
\nonumber \\ &&
=\sum_{nmk}\frac{2 n_c g^2_{nmk0}M}{\hbar^2}
\sqrt{|p_{cz}-p_{z2}|^2-4M(U^{n}+E_n-\mu_c)}
\nonumber \\ &&
\times (1+f_3^m)(1+f_4^k)\Delta t 
\nonumber \\ &&
=\sum_{k}\frac{2 n_c g^2_{nmk0}M}{\hbar^2}
\sqrt{|p_{cz}-p_{z2}|^2-4M(E_n-E_0+g'_{nn00}n_c)}
\nonumber \\ &&
\times (1+f_3^m)(1+f_4^k)\Delta t ,
\label{P^{out}_{12}}
\end{eqnarray}
where we assume 
\begin{eqnarray}
U^{n}-\mu_c&=&V_{\rm ext}(z)+2(g_{nn00}n_c+\sum_{k}g_{nnkk}\tilde n^{k})
-(V_{\rm ext}(z)+g_{0000}n_c+\sum_{k}g_{00kk}\tilde n^{k})
\nonumber \\ &\simeq&\sum_{k}(2g_{nnkk}-g_{00kk})n_c
\equiv g'_{nn00}n_c.
\end{eqnarray}
The ``in" collision rate is given by
\begin{eqnarray}
\Gamma_{12}^{in}[f_{n}]&=&\sum_{n'm'k'}\frac{g_{n'm'k'0}^2n_c}{\pi\hbar^3}
 \int dp_{z2} \int dp_{z3}\int dp_{z4} \delta(\epsilon_c+\tilde{\epsilon}_2^{n'}-\tilde{\epsilon}_3^{m'}-\tilde{\epsilon}_4^{k'}) (\delta_{nn'}-\delta_{nm'}-\delta_{nk'})
\nonumber \\ && \times \delta({p_{cz}+p_{z2}-p_{z3}-p_{z4}}) (1+f_2^{n'})f_3^{m'}f_4^{k'}
\nonumber \\ &=&
\sum_{n'm'k'} \int \frac{ dp_{z2}}{2\pi \hbar}f_2^{m'}\int \frac{ dp_{z4}}{2\pi \hbar}f_4^{k'}\frac{g_{n'm'k'0}^2Mn_c}{\pi\hbar}(1+f_3^{n'})(\delta_{nn'}-\delta_{nm'}-\delta_{nk'})
\nonumber \\ &&
\delta[(p_{cz}-p_{z4})(p_{cz}-p_{z2})-Mg'_{n'n'00}n_c+M(E_0+E_{n'}-E_{m'}-E_{k'})],
\nonumber \\ &&
\end{eqnarray}
where we have interchanged the particle labels 2 and 3 to
obtain the second line in this equation. This rate corresponds
to two thermal atoms scattering into a condensate atom and
an outgoing thermal atom, and is thus the rate that atoms
feed into the condensate as a result of collisions.
Although
the collision of atoms 2 and 4 can be treated by the methods
used to analyze the $C_{22}$ collision rate, it is preferable to define
a single atom collision rate by writing this integral in the
form of Eq. (\ref{P^{out}_{12}}) and performing a Monte Carlo sampling
with respect to the $p_{z2}$ variable. This procedure leads to the
collision probability per atom
\begin{eqnarray}
P_{i_{nmk}}^{in}[f_{n}]&=&\Delta t \int \frac{ dp_{z4}}{2\pi \hbar}f_4^{k}\frac{g_{nmk0}^2Mn_c}{\pi\hbar}(1+f_3^{n})
\nonumber \\ &&
\delta[(p_{cz}-p_{z4})(p_{cz}-p_{z2}^i)-Mg'_{nn00}n_c+M(E_0+E_{n}-E_{m}-E_{k})].
\nonumber \\ &&
\end{eqnarray}
This analysis yields probabilities for a particular atom to
undergo `out' or `in' collisions. To decide whether either
event takes place, another random number $0<X_{12}<1$ is chosen.
If $X_{12}<P_i^{\rm out}$ then an `out' collision is accepted; the
incoming thermal atom is removed from the ensemble of test
particles and two new thermal atoms are created. However, if
$P_i^{\rm out}<X_{12}<P_i^{\rm out}+P_i^{ \rm in}$ , then an `in' collision takes place and atom 2 is removed from the thermal sample. In addition, a
second test particle, atom 4, is removed and a new thermal
atom, atom 3, is created. In practice, it is exceedingly unlikely
that a test particle will exist that will precisely match
the required phase-space coordinates of particle 4. We therefore
search for a test particle in neighboring phase-space
cells and remove this particle if one is found. This can be
justified by remembering that we are only interested in describing
the evolution in phase-space in a statistical way\it
is misleading to think of a direct correspondence between the
test particles and physical atoms. If no test particle exists in
the vicinity of $v_4$, the local phase-space density $f_4$, and
hence $P_i^{ \rm in}$, will be zero and the `in' collision is precluded
from occurring in any case.
The above procedure leads to a change in the number of
atoms in the thermal cloud. In order to conserve the total
particle number the GP equation is propagated with the R
term which changes the normalization of the wave function
and hence the condensate number. This quantity can be
evaluated from the Monte Carlo process decribed above by
summing probabilities for particles
\begin{eqnarray}
R(z,t)=\frac{\hbar}{2n_c\Delta t}\sum_{nmk}\sum_i(P_{i_{nmk}}^{\rm out}-P_{i_{nmk}}^{\rm in}).
\end{eqnarray}
In practice, this assignment to grid points is performed with
a cloud-in-cell approach similar to the one described earlier.
Of course, the normalization of the condensate wave function
varies continuously as opposed to the variation of the
thermal atom number which changes by discrete jumps. Nevertheless,
one can show that the subsequent change in the
condensate normalization is consistent with the addition or
removal of atoms from the thermal cloud, so that the total
particle number, $N_{\rm tot}$ , is conserved within statistical fluctuations.


\begin{thebibliography}{20}
\expandafter\ifx\csname natexlab\endcsname\relax\def\natexlab#1{#1}\fi
\expandafter\ifx\csname bibnamefont\endcsname\relax
  \def\bibnamefont#1{#1}\fi
\expandafter\ifx\csname bibfnamefont\endcsname\relax
  \def\bibfnamefont#1{#1}\fi
\expandafter\ifx\csname citenamefont\endcsname\relax
  \def\citenamefont#1{#1}\fi
\expandafter\ifx\csname url\endcsname\relax
  \def\url#1{\texttt{#1}}\fi
\expandafter\ifx\csname urlprefix\endcsname\relax\def\urlprefix{URL }\fi
\providecommand{\bibinfo}[2]{#2}
\providecommand{\eprint}[2][]{\url{#2}}

\bibitem[{\citenamefont{Heiselberg}(2006)}]{T_Heiselberg_Unitarysound}
\bibinfo{author}{\bibfnamefont{H.}~\bibnamefont{Heiselberg}},
  \bibinfo{journal}{Phys. Rev. A} \textbf{\bibinfo{volume}{73}},
  \bibinfo{pages}{013607} (\bibinfo{year}{2006}).

\bibitem[{\citenamefont{Capuzzi et~al.}(2006)\citenamefont{Capuzzi, Vignolo,
  Federici, and Tosi}}]{T_C}
\bibinfo{author}{\bibfnamefont{P.}~\bibnamefont{Capuzzi}},
  \bibinfo{author}{\bibfnamefont{P.}~\bibnamefont{Vignolo}},
  \bibinfo{author}{\bibfnamefont{F.}~\bibnamefont{Federici}}, \bibnamefont{and}
  \bibinfo{author}{\bibfnamefont{M.~P.} \bibnamefont{Tosi}},
  \bibinfo{journal}{Phys. Rev. A} \textbf{\bibinfo{volume}{73}},
  \bibinfo{pages}{021603(R)} (\bibinfo{year}{2006}).

\bibitem[{\citenamefont{Joseph et~al.}(2007)\citenamefont{Joseph, Clancy, Luo,
  Kinast, Turlapov, and Thomas}}]{E_Joseph_PRL98}
\bibinfo{author}{\bibfnamefont{J.}~\bibnamefont{Joseph}},
  \bibinfo{author}{\bibfnamefont{B.}~\bibnamefont{Clancy}},
  \bibinfo{author}{\bibfnamefont{L.}~\bibnamefont{Luo}},
  \bibinfo{author}{\bibfnamefont{J.}~\bibnamefont{Kinast}},
  \bibinfo{author}{\bibfnamefont{A.}~\bibnamefont{Turlapov}}, \bibnamefont{and}
  \bibinfo{author}{\bibfnamefont{J.~E.} \bibnamefont{Thomas}},
  \bibinfo{journal}{Phys. Rev. Lett.} \textbf{\bibinfo{volume}{98}},
  \bibinfo{pages}{170401} (\bibinfo{year}{2007}).

\bibitem[{\citenamefont{Meppelink et~al.}(2009)\citenamefont{Meppelink, Koller,
  and van~der Straten}}]{E_second_Bose}
\bibinfo{author}{\bibfnamefont{R.}~\bibnamefont{Meppelink}},
  \bibinfo{author}{\bibfnamefont{S.~B.} \bibnamefont{Koller}},
  \bibnamefont{and} \bibinfo{author}{\bibfnamefont{P.}~\bibnamefont{van~der
  Straten}}, \bibinfo{journal}{Phys. Rev. A} \textbf{\bibinfo{volume}{80}},
  \bibinfo{pages}{043605} (\bibinfo{year}{2009}).

\bibitem[{\citenamefont{Hu et~al.}(2010)\citenamefont{Hu, Taylor, Liu,
  Stringari, and Griffin}}]{E_second_BF}
\bibinfo{author}{\bibfnamefont{H.}~\bibnamefont{Hu}},
  \bibinfo{author}{\bibfnamefont{E.}~\bibnamefont{Taylor}},
  \bibinfo{author}{\bibfnamefont{X.-J.} \bibnamefont{Liu}},
  \bibinfo{author}{\bibfnamefont{S.}~\bibnamefont{Stringari}},
  \bibnamefont{and} \bibinfo{author}{\bibfnamefont{A.}~\bibnamefont{Griffin}},
  \bibinfo{journal}{New J. Phys.} \textbf{\bibinfo{volume}{12}},
  \bibinfo{pages}{043040} (\bibinfo{year}{2010}).

\bibitem[{\citenamefont{Griffin et~al.}(2009)\citenamefont{Griffin, Nikun, and
  Zaremba}}]{B_GNZ}
\bibinfo{author}{\bibfnamefont{A.}~\bibnamefont{Griffin}},
  \bibinfo{author}{\bibfnamefont{T.}~\bibnamefont{Nikun}}, \bibnamefont{and}
  \bibinfo{author}{\bibfnamefont{E.}~\bibnamefont{Zaremba}},
  \emph{\bibinfo{title}{Bose-Condensed Gases at Finite Temperatures}}
  (\bibinfo{publisher}{UNIVERSITY PRESS CAMBRIDGE}, \bibinfo{year}{2009}).

\bibitem[{\citenamefont{Landau}(1941)}]{Hydro_Landau}
\bibinfo{author}{\bibfnamefont{L.~D.} \bibnamefont{Landau}},
  \bibinfo{journal}{J. Phys. (USSR)} \textbf{\bibinfo{volume}{5}},
  \bibinfo{pages}{71} (\bibinfo{year}{1941}).

\bibitem[{\citenamefont{Kinast et~al.}(2004)\citenamefont{Kinast, Hemmer, Gehm,
  Turlapov, and Thomas}}]{E_Kinast_PRL92}
\bibinfo{author}{\bibfnamefont{J.}~\bibnamefont{Kinast}},
  \bibinfo{author}{\bibfnamefont{S.~L.} \bibnamefont{Hemmer}},
  \bibinfo{author}{\bibfnamefont{M.~E.} \bibnamefont{Gehm}},
  \bibinfo{author}{\bibfnamefont{A.}~\bibnamefont{Turlapov}}, \bibnamefont{and}
  \bibinfo{author}{\bibfnamefont{J.~E.} \bibnamefont{Thomas}},
  \bibinfo{journal}{Phys. Rev. Lett.} \textbf{\bibinfo{volume}{92}},
  \bibinfo{pages}{150402} (\bibinfo{year}{2004}).

\bibitem[{\citenamefont{Bartenstein et~al.}(2004)\citenamefont{Bartenstein,
  Altmeyer, Riedl, Jochim, Chin, Denschlag, and Grimm}}]{E_Bartenstein_PRL92}
\bibinfo{author}{\bibfnamefont{M.}~\bibnamefont{Bartenstein}},
  \bibinfo{author}{\bibfnamefont{A.}~\bibnamefont{Altmeyer}},
  \bibinfo{author}{\bibfnamefont{S.}~\bibnamefont{Riedl}},
  \bibinfo{author}{\bibfnamefont{S.}~\bibnamefont{Jochim}},
  \bibinfo{author}{\bibfnamefont{C.}~\bibnamefont{Chin}},
  \bibinfo{author}{\bibfnamefont{J.~H.} \bibnamefont{Denschlag}},
  \bibnamefont{and} \bibinfo{author}{\bibfnamefont{R.}~\bibnamefont{Grimm}},
  \bibinfo{journal}{Phys. Rev. Lett} \textbf{\bibinfo{volume}{92}},
  \bibinfo{pages}{203201} (\bibinfo{year}{2004}).

\bibitem[{\citenamefont{Andrews et~al.}(1997)\citenamefont{Andrews, Kurn,
  Miesner, Durfee, Townsend, Inouye, and Ketterle}}]{E_Andrews_PRL79}
\bibinfo{author}{\bibfnamefont{M.~R.} \bibnamefont{Andrews}},
  \bibinfo{author}{\bibfnamefont{D.~M.} \bibnamefont{Kurn}},
  \bibinfo{author}{\bibfnamefont{H.-J.} \bibnamefont{Miesner}},
  \bibinfo{author}{\bibfnamefont{D.~S.} \bibnamefont{Durfee}},
  \bibinfo{author}{\bibfnamefont{C.~G.} \bibnamefont{Townsend}},
  \bibinfo{author}{\bibfnamefont{S.}~\bibnamefont{Inouye}}, \bibnamefont{and}
  \bibinfo{author}{\bibfnamefont{W.}~\bibnamefont{Ketterle}},
  \bibinfo{journal}{Phys. Rev. Lett} \textbf{\bibinfo{volume}{79}},
  \bibinfo{pages}{553} (\bibinfo{year}{1997}).

\bibitem[{\citenamefont{Stringari}(1996)}]{T_stringari_PRL}
\bibinfo{author}{\bibfnamefont{S.}~\bibnamefont{Stringari}},
  \bibinfo{journal}{Phys. Rev. Lett.} \textbf{\bibinfo{volume}{2360}},
  \bibinfo{pages}{77} (\bibinfo{year}{1996}).

\bibitem[{\citenamefont{Griffin and Zaremba}(1997)}]{T_Griffin_PRA56}
\bibinfo{author}{\bibfnamefont{A.}~\bibnamefont{Griffin}} \bibnamefont{and}
  \bibinfo{author}{\bibfnamefont{E.}~\bibnamefont{Zaremba}},
  \bibinfo{journal}{Phys. Rev. A} \textbf{\bibinfo{volume}{56}},
  \bibinfo{pages}{4839} (\bibinfo{year}{1997}).

\bibitem[{\citenamefont{Jackson et~al.}(2007)\citenamefont{Jackson, Proukakis,
  and Barenghi}}]{T_Jacson_PRA75}
\bibinfo{author}{\bibfnamefont{B.}~\bibnamefont{Jackson}},
  \bibinfo{author}{\bibfnamefont{N.~P.} \bibnamefont{Proukakis}},
  \bibnamefont{and} \bibinfo{author}{\bibfnamefont{C.~F.}
  \bibnamefont{Barenghi}}, \bibinfo{journal}{Phys. Rev. A}
  \textbf{\bibinfo{volume}{75}}, \bibinfo{pages}{051601(R)}
  (\bibinfo{year}{2007}).

\bibitem[{\citenamefont{Arahata et~al.}(2011)\citenamefont{Arahata, Nikuni, and
  Griffin}}]{T_Arahata_PRA2011}
\bibinfo{author}{\bibfnamefont{E.}~\bibnamefont{Arahata}},
  \bibinfo{author}{\bibfnamefont{T.}~\bibnamefont{Nikuni}}, \bibnamefont{and}
  \bibinfo{author}{\bibfnamefont{A.}~\bibnamefont{Griffin}},
  \bibinfo{journal}{Phys. Rev. A} \textbf{\bibinfo{volume}{84}},
  \bibinfo{pages}{053612} (\bibinfo{year}{2011}).

\bibitem[{\citenamefont{Nikuni and Griffin}(2001)}]{T_tau_Nikuni}
\bibinfo{author}{\bibfnamefont{T.}~\bibnamefont{Nikuni}} \bibnamefont{and}
  \bibinfo{author}{\bibfnamefont{A.}~\bibnamefont{Griffin}},
  \bibinfo{journal}{Phys. Rev. A} \textbf{\bibinfo{volume}{65}},
  \bibinfo{pages}{011601} (\bibinfo{year}{2001}).

\bibitem[{\citenamefont{Arahata and Nikuni}(2009)}]{T_Ara_PRA2009_F}
\bibinfo{author}{\bibfnamefont{E.}~\bibnamefont{Arahata}} \bibnamefont{and}
  \bibinfo{author}{\bibfnamefont{T.}~\bibnamefont{Nikuni}},
  \bibinfo{journal}{Phys. Rev. A} \textbf{\bibinfo{volume}{80}},
  \bibinfo{pages}{043613} (\bibinfo{year}{2009}).

\bibitem[{\citenamefont{Zaremba et~al.}(2009)\citenamefont{Zaremba, Nikuni, and
  Griffin}}]{T_ZNG_JLTP}
\bibinfo{author}{\bibfnamefont{E.}~\bibnamefont{Zaremba}},
  \bibinfo{author}{\bibfnamefont{T.}~\bibnamefont{Nikuni}}, \bibnamefont{and}
  \bibinfo{author}{\bibfnamefont{A.}~\bibnamefont{Griffin}},
  \bibinfo{journal}{J. Low Temp. Phys.} \textbf{\bibinfo{volume}{116}},
  \bibinfo{pages}{277} (\bibinfo{year}{2009}).

\bibitem[{\citenamefont{Arahata and Nikuni}(2008)}]{T_Ara_PRA2008}
\bibinfo{author}{\bibfnamefont{E.}~\bibnamefont{Arahata}} \bibnamefont{and}
  \bibinfo{author}{\bibfnamefont{T.}~\bibnamefont{Nikuni}},
  \bibinfo{journal}{Phys. Rev. A} \textbf{\bibinfo{volume}{77}},
  \bibinfo{pages}{033610} (\bibinfo{year}{2008}).

\bibitem[{\citenamefont{Jackson and Zaremba}(2002)}]{T_Jacson_Zaremba}
\bibinfo{author}{\bibfnamefont{B.}~\bibnamefont{Jackson}} \bibnamefont{and}
  \bibinfo{author}{\bibfnamefont{E.}~\bibnamefont{Zaremba}},
  \bibinfo{journal}{Phys. Rev. A} \textbf{\bibinfo{volume}{66}},
  \bibinfo{pages}{033606} (\bibinfo{year}{2002}).

\bibitem[{\citenamefont{Zaremba et~al.}(1998)\citenamefont{Zaremba, Griffin,
  and Nikuni}}]{T_ZNG_sound}
\bibinfo{author}{\bibfnamefont{E.}~\bibnamefont{Zaremba}},
  \bibinfo{author}{\bibfnamefont{A.}~\bibnamefont{Griffin}}, \bibnamefont{and}
  \bibinfo{author}{\bibfnamefont{T.}~\bibnamefont{Nikuni}},
  \bibinfo{journal}{Phys. Rev. A} \textbf{\bibinfo{volume}{57}},
  \bibinfo{pages}{4695} (\bibinfo{year}{1998}).

\end{thebibliography}
\end{document}